# NATURAL CALCIUM CARBONATE FOR BIOMEDICAL APPLICATIONS

A thesis submitted in partial fulfilment for the degree of

MASTER OF TECHNOLOGY (INTEGRATED)

IN

BIOTECHNOLOGY

*Submitted by*

SONALI SUDHIR SALI

Under the Guidance of

Dr. S. VIJAYALAKSHMI

CENTRE FOR RESEARCH IN NANOTECHNOLOGY AND SCIENCE

INDIAN INSTITUTE OF TECHNOLOGY – BOMBAY (IITB)

Powai, Mumbai – 400076.

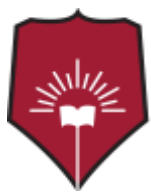
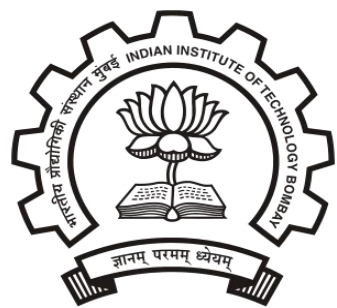

SCHOOL OF BIOTECHNOLOGY AND BIOINFORMATICS

D. Y. PATIL UNIVERSITY

MAY – 2015

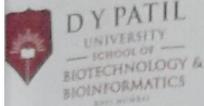

**D. Y. PATIL UNIVERSITY, NAVI MUMBAI**
ACCREDITED BY NAAC WITH 'A' GRADE
**SCHOOL OF BIOTECHNOLOGY & BIOINFORMATICS**

Dr. Debjani Dasgupta
Director

# CERTIFICATE

This is to certify that **Ms. Sali Sonali Sudhir Vaishali**, M. Tech Integrated (X Semester) of the School of Biotechnology and Bioinformatics, carried out the Dissertation entitled, **"NATURAL CALCIUM CARBONATE FOR BIOMEDICAL APPLICATIONS"** for the partial fulfillment of Master of Technology (Integrated) in Biotechnology. The dissertation has not formed the basis for the award of any degree, diploma, associate-ship or fellowship. The dissertation represents independent work carried out by the candidate.

Place : Navi Mumbai

Date: 21st May 2015

*Debjani D*

**Dr. Debjani Dasgupta**
**Director**

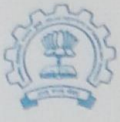

नैनो प्रौद्योगिकी तथा विज्ञान अनुसंधान
भारतीय प्रौद्योगिकी संस्थान मुंबई
पवई, मुंबई-400 076, भारत
**Centre for Research in Nano Technology and Science**
Indian Institute of Technology Bombay
Powai, Mumbai-400 076, India

Tel : (+91-22) 2576 7691, 2576 7692,
(+91-22) 2576 7690, 2572 3806 (DIR)
EPABX : (+91-22) 2572 2545,
Extn.: 7691, 7692, 7690
Fax : (+91-22) 2572 3480, 2572 3314 (DIR)
Website : www.iitb.ac.in

# CERTIFICATE

I certify that the research work presented in this thesis titled "Natural Calcium carbonate for biomedical applications" has been carried out by Ms. Sonali Sudhir Sali, Roll No. MTI – 10052 under my supervision and this is her bonafide work. The research work is original and has not been submitted for any other University. Further she was a regular student and has worked under my guidance as a full time student at Indian Institute of Technology Bombay (IITB) until the submission of the thesis to the D.Y.Patil University.

Date: 29-05-2015
Place: IIT-B, POWAI

Guide's Signature
(Dr. S. Vijayalakshmi)
Research Scientist, CRNTS

# DECLARATION BY THE CANDIDATE

This is to state that the work embodied in this thesis titled "Natural Calcium carbonate for biomedical applications" forms my own contribution to the research work carried out under the guidance of Dr. S. Vijayalakshmi at Indian Institute of Technology Bombay (IITB). This work has not been submitted for any degree for this University or any other University. Whenever references have been made to previous work of others, it has been clearly indicated as such and included in the Bibliography.

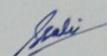

Signature of candidate

(SonaliSudhirSali)

Roll No.: MTI - 10052

***Dedicated to my mother…***

# ACKNOWLEDGEMENTS

This thesis would simply not have been possible without the work, guidance and endless belief and support of my guide **Dr. S. Vijayalakshmi**. She really did show me what it takes to pursue research and kept me going. Her patient, modest and kind nature helped me understand and do better. Her warm words of appreciation and encouragement made me want to go ahead in the field, which will never be forgotten.

I would also like to thank **Mohammad Furquan** who helped and supported throughout the project.

I extend gratitude to my fellow interns and friends for their genuine encouragement; **Hafis Rahman** and **Sakshi Anand** and **Varun S.**, discussions with whom proved very useful.

And lastly I am grateful to my mother **Vaishali N. Sarode** for her constant love and support throughout difficult times.

# ABSTRACT

Sea shells are found to be a very rich natural resource for calcium carbonate. Nanoparticles found in sea shells are in the prime target of researchers in the biomedical application field. Sea shells are made up of $CaCO_3$ mainly in the aragonite form, which are columnar or fibrous or microsphere structured crystals. The porous nature of $CaCO_3$ is advantageous for various drug delivery applications. The bioactivity of nanoparticles of sea shell has been studied in this work**.** The sea shells collected from the beach were thoroughly washed, dried and pulverized. The powder was sieved and particles in the range of 45-63 microns were collected. The powdered sea shells were characterized using X-Ray Diffraction (XRD) and Field Emission Gun-Scanning Electron Microscopy (FE-SEM). The XRD data showed that the particles were mainly microspheres. Traces of calcite and vaterite were also present. Experiments were conducted to study the aspirin and strontium ranelate drug loading into the sea shell powder using soak and dry method. Different concentration of (500 µg to 10 mg/ml) drug solution was made in ethanol and water. About 100 mg of the sea shell powder was soaked in 5 ml of the drug solutions and was kept soaking for 48 hrs with intermittent ultrasonication. The mixture was then gently dried in a vacuum oven at $40^oC$. The FE-SEM images were taken for all the feed and product samples. The *in vitro* drug release studies were done using Phosphate Buffered Saline (PBS). The FE-SEM images displayed a distribution of differently sized and shaped particles. The sea shells due to its natural porosity and crystallinity are expected to be useful for drug delivery. About 50% drug entrapment efficiency for aspirin and 39% for strontium ranelate was seen. A burst release of the drug (80%) was observed within two hours for both the drugs studied. Rest of the drug was released slowly in 19 hrs. Further modification of the sea shell with non-toxic polymers is also planned as a part of this work. Sea shell powder has become a potential candidate for drug delivery due to all the aforementioned advantages.

# TABLE OF CONTENTS

## CHAPTER 3

# CHAPTER 4

# LIST OF FIGURES

# LIST OF TABLES

# ABBREVIATIONS

**CaCO$_3$**   –   Calcium Carbonate

**XRD**   –   X-Ray Diffraction

**FE-SEM**   –   Field Emission-Scanning Electron Microscopy

**TEM**   –   Transmission Electron Microscopy

**HR-LCMS**   –   High Resolution-Liquid Chromatography Mass Spectrometry

**PBS**   –   Phosphate Buffer Saline

**SR**   –   Strontium ranelate

**EDS**   –   Energy Dispersive X-Ray Spectroscopy

**FTIR**   –   Fourier Transform Infrared Spectroscopy

**TGA**   –   Thermogravimetric Analysis

**ASA**   –   Acetylsalysilic acid

# CHAPTER 1

## 1.1 INTRODUCTION

## 1.1.1  BIOMEDICAL APPLICATIONS OF NANOTECHNOLOGY

There are three main categories of applications in biomedicine: drugs, diagnostic techniques, and prostheses and implants. There is a lot of interest booming for applications outside the body in diagnostic **biosensors** and 'lab-on-a-chip' techniques to analyse blood and other samples. The biological sensors or biosensors are made up of a biological element that may be an antibody, an enzyme or a nucleic acid, and a transducer. The biological element fixed in the chip interacts with the analyte being tested which creates a biological response is efficiently converted to an electrical signal by the transducer. Biological sensors have become extremely valuable devices to measure a broad range of spectra of analytes comprising of gases, ions, organic compounds and bacteria.

**Nanodrugs** is a concept that is on an exponential rise today in nanotechnology research. Be it carbon buckyball or nanotubes or any other nanopowder, they contribute generously to the drug delivery system. Nanodrugs are designed to greatly enhance their capacity to conduct therapeutics. At first the nanodrugs comprised of anti-cancer drugs or any other drugs loaded into/onto synthesized nanocarriers without any targeting features. Non-targeted nanodrugs have given an opportunity to carry large amounts of drugs, along with poorly water-soluble and/or permeable drugs. Many products like toothpastes, sunscreens and various medicines are already commercialised and sold in the market today. Quantum dot based cosmetics are sold in large quantities. [1]

Clinical diagnostics is the most promising application for very little amounts of fluids that can be tested on lab-on-a-chip system. Conventional or traditional **microfluidic devices** are based on continuous unbroken fluid film flow in microchannels, but give very little flexibility in reconfigurability and scalability. Thus a fully integrated and reconfigurable "digital" microfluidic lab-on-a-chip which is droplet based is made for clinical diagnosis of human physiological fluids has been designed. The droplets act as solution-phase in the reaction chambers on the chip and are manipulated by using the electrowetting effect. Thus we can obtain a repeatable and reliable high-speed movement of microdroplets of human whole blood, plasma, saliva, serum, sweat, tear and urine which is demonstrated to attain a good compatibility of these fluids with the chip's electrowetting platform. The lab-on-a-chip comprises of sample injection on chip reservoirs, droplet forming structures, fluid paving its pathways, mixing areas wherever necessary and lastly optical detection sites.

Another application is **imaging** for therapy in which there are sophisticated machines used to obtain images of the inside of your body. Molecular imaging is not merely a traditional process of forming an image and interpreting it, but it is intended for diagnostic accuracy and sensitivity by giving an in vivo image of cell-drug chemistry or interaction and effects. It gives an efficient image contrast and high definition resolution which are necessary to view images properly, but it focuses more on depicting the enhanced consequences of microscopic pathologies by focusing on the molecular components or processes that show the actual mechanisms of diseases. Medical imaging technologies help for instant diagnosis and analysis of various pathologies. To increase the sensitivity along with usage, technologies like CT and MRI depend on administering contrast agents intravenously. The current types of contrast agents have definitely allowed for rapid diagnosis but they still have many negative points including no tissue specificity and systemic toxicity. Through the numerous advances made in nanotechnology and materials science, now creation a new generation of contrast agents that eliminate these problems, and are capable of providing more sensitive and specific information. [1] [2] [3] [4]

Nanotechnology is currently under intense development for its applications in cancer imaging, molecular diagnosis and targeted drug delivery. Cancer was previously considered an incurable disease. Today most of the patients diagnosed at an early stage survive their illness. Advances in cancer diagnostics and therapeutics in the last few years are majorly responsible for this dramatic improvement. Delivering drug-loaded nanoparticles to cancerous cells has been focused on passive and active targeting methods. In comparison with passive targeting, that uses manipulation of kinetics of the drugs and size reduction of nanoparticles, active targeting is done by delivering drug-loaded nanoparticles to specifically identified sites while having minimum undesirable effect anywhere else in the body. Active cancer cell targeting is done by administration of nanoparticles with targeting molecules bound on the particle surface that can detect or recognize and then bind to specific ligands that in turn recognize only the cancer cells. In local drug delivery, the cytotoxic drug loaded in the nanoparticles can be directly delivered to cancerous cells so that there is minimum harmful toxicity of the drug towards the non-cancerous cells surrounding the targeted tissue. This approach is useful for initial primary tumours that have not yet become metastatic in nature. For metastatic cancer cells, the location, size and amount of tumour in the body must be visualized or accessed precisely, which is not exactly possible due to tumour nature thus making local delivery impractical. In this case, the drug delivery vehicle would be administered systemically, in which the drug travels through the entire body. An active compound might be bonded to a particle's surface or inserted in a nanotube. When linked with ligands that target biological compounds, such as peptides, monoclonal antibodies or small molecules, the nanoparticles are efficiently used to target malignant cancerous tumours with high binding capacity and specificity. In the diameter size range of 5-100 nm, nanoparticles have a large surface area and a functional group for binding to multiple diagnostic and therapeutic agents. Recent advances have thus led to multifunctional nanoparticle for targeted cancer therapy. [5] [6]

## 1.1.2 NANOTECHNOLOGY BASED DRUG DELIVERY SYSTEMS

Conventional drug administration forms are pills, ointments, eye drops, and intravenous solutions. Today, various new drug delivery systems have been developed. These systems include drug modification chemically, drug entrapment in polymeric materials, drug entrapment in small pores that are injected later into the bloodstream. This system is the absorption of drug across a biological membrane, but the targeted release system releases it in the form of a dosage. The plus point of the targeted release system is the reduction in the periodic outflow of the dosages, thus having a uniform effect of the drug, reduction of its side-effects, and also reduced irregularity in circulating drug levels. The disadvantage is high cost, making productivity difficult and lesser ability of adjusting the dosages. Thus only a certain amount of a therapeutic agent is released for a long time period into a targeted diseased area. Thus prevents damage to the healthy tissue.

Nano drug delivery takes advantage that, nanoscaled materials ($10^{-9}$ to $10^{-7}$ m) can possess very different physical properties, mechanical and optical and electrical which are different from those seen in the macroscopic counterparts. Nano-scale drug-delivery systems can be produced to mix and match different biological and synthetic modules, for various applications, comprising of injectable, oral, topical, implantable, inhalable and transdermal drug delivery. Nanoparticles can also be functionalized with biomolecules by various methods including physical adsorption, binding recognition and covalent coupling. Nanoparticles can be modified to get efficient targeting to specific organelles. Many of the properties of these delivery systems can be tailored desirably for specific applications such as solubility, biodistribution, biocompatibility, biodegradability, drug release, drug encapsulation and shape. Spherical nanoparticles are the simplest to create but other shapes and constructions also offer advantages for certain applications. [7]

The various materials are polymers, nanocapsules, nanotubes and nanogels. To look into a few in depth, firstly **polymers.** These types of therapeutics include specially designed drugs that can be macromolecules. The advantage is their comfort with chemical changes, giving a good chemical composition, functional surface and the possibility of three-dimensional structures. Several polymers are used for clinical therapies: synthetic polymers, natural polymers and pseudosynthetic polymers. Polymers offer diversity in chemistry, varied possible morphologies, allowing them a good use of materials that are suitable for applications in nanoscale drug-delivery systems. Study of the structural and functional relationships of polymers is increasing their utility. The range of polymeric structures is quite wide: linear, block, graft, multivalent, branched, cross-linked, dendronic and star-shaped polymers. Polymer structure can be of great importance as the effectiveness of drug carriers depends on the polymer's chemical composition, backbone stability and water solubility. The polymer structure not only affects the carrier's physicochemical properties but also affects its drug-loading efficacy, rate of drug-release and also rendering drastic effect on its bioavailability. Drugs can be physically loaded in the polymers or covalently attached to the polymer backbone. Polymers are heterogeneous mixtures of chains of different lengths. Recent research has shown that polymers are suited for *in vivo* applications. Scientists are meanwhile continuing to explore new biodegradable polymers that show better three-dimensional structures and are more efficiently suited for frequent parenteral administration.

Lipid and polymeric **nanocapsules** can provide controlled release of the drug and efficient site targeting. The composition of the outer coating with this material, determines their dispersion stability. Nanocapsule fabrication can be done by precipitation, layer-by-layer deposition and self-assembly procedures. Important parameters are capsule size and radius distribution, thickness, membrane quality and type. Lipid-based nanocapsules can be easily modified artificially to bring desired changes in the membrane permeability by inserting channel and targeting specific cells by attaching antibodies. The use of lipids is although limited because of their instability inside the biological media and also because of their sensitivity to a wide range of external parameters, like temperature, pH and osmotic pressure. The stability these nanocapsules can be improved by conjugating lipids with polymers.

**Nanotubes** offer advantages more than spherical nanoparticles for some applications. The inner volumes can be filled with drugs and because the inner and outer surfaces of some types of nanotubes are different, they are separately modified to take in the drugs. Finally, the open-mouthed structure on both the sides makes drug loading easy. Nanotubes can be fabricated from many materials and *via* distinct routes, ranging from self-assembly to deliberate deposition. Examples include fullerene carbon nanotubes, cyclic peptide nanotubes and template-synthesized nanotubes. In template approach that is the most versatile way to fabricate nanotubes in which they are made by depositing the material (polymer, silica, metal, or carbon) within the cylindrical pores of a solid surface. The outer diameter is decided by the template diameter and the inner diameter by its deposition time. Future work may focus on manipulating the coating to control the drug release. This approach has several advantages like increased control, greater efficiency than standard transfection reagents and decreased cytotoxicity.

**Hydrogel** matrices are biocompatible and are used in drug delivery as they can prevent payload aggregation. Hydrogels as drug carriers have the biggest advantage that they can be synthesized in the absence of drugs. Generally, the drug is loaded along by self-assembly based on non-covalent interactions between them. Nano-scale hydrogels, or "nanogels", offer a very high drug-loading capacity. Hydrogels are hydrophilic in nature, have three-dimensional cross-linked networks that swell up when in contact with water. They respond to ionic strength, pH and temperature. They combine the properties of gels with those of colloids, which are high surface-to-volume ratio and small size. Polymeric nanogels have long stability, controlled release, low levels of cytotoxicity and have better protection from enzymatic degradation. However hydrogel coating is difficult although they have been coated with lipids to manage colloidal stability, but they have low coating efficiency.

**Stem cells** have a huge potential in medicine: from repairing heart cells to replacing nerve cells which are lost in the brain of a patient with Parkinson's disease. Using stem cells as a therapy means making them grow into the desired type of tissue. Inside the body this happens due to exact chemical and physical signals sent by brain, but not all of them are understood or characterised. Usage of chemicals to make the stem cells to desired cells has worked in laboratories, but results are not often safe or predictable. A team from North-western University in the US has said they have a solution according to which they can direct the development of stem cells using physical properties, by transforming a technique called scanning probe lithography that traces 3-D microscopic shapes and constructs them on flat surfaces. Placing the stem cells on this surface, devoid of any chemicals, the stem cells can be induced to develop into bone cells. Extending this same technique, it may be possible to transform stem cells into any type of cells. The body needs repairs to be carried out for which mesenchymal stem cells or MSCs can enter the blood circulation system and travel around the body and get attached wherever they are needed. MSCs can develop into a wide range of tissue types, so they are pluripotent. The developments that happen depend partly, on the molecular structures in the matrix around the cells which make up the tissue. Scientists have mimicked this real-life situation, by using the molecular structures in the matrix around the cell as a pattern. Then with an array of pyramid-like points that are thousand times smaller than the tip of a pencil are used to build a nanolandscape, molecule by molecule, with sculptures in sizes ranging from nano to microscale, on a small glass piece. This technique is called as polymer pen lithography. The researchers grew MSCs on a type of nanoscopic sculpture, and also were able to direct the stem cells into osteocytes (bone cells). The purpose and potential of this tool is to take pluripotent stem cells from a patient, grow them on a selected 3-D matrix to convert them progressively and rapidly into any particular type of cell of our choice. Then return the cells to the patient for repair of damaged tissues. One important aspect of this work is that, it provides proof that stem-cell developmental fate can be changed by the solely using the 3-D molecular structure. [8]

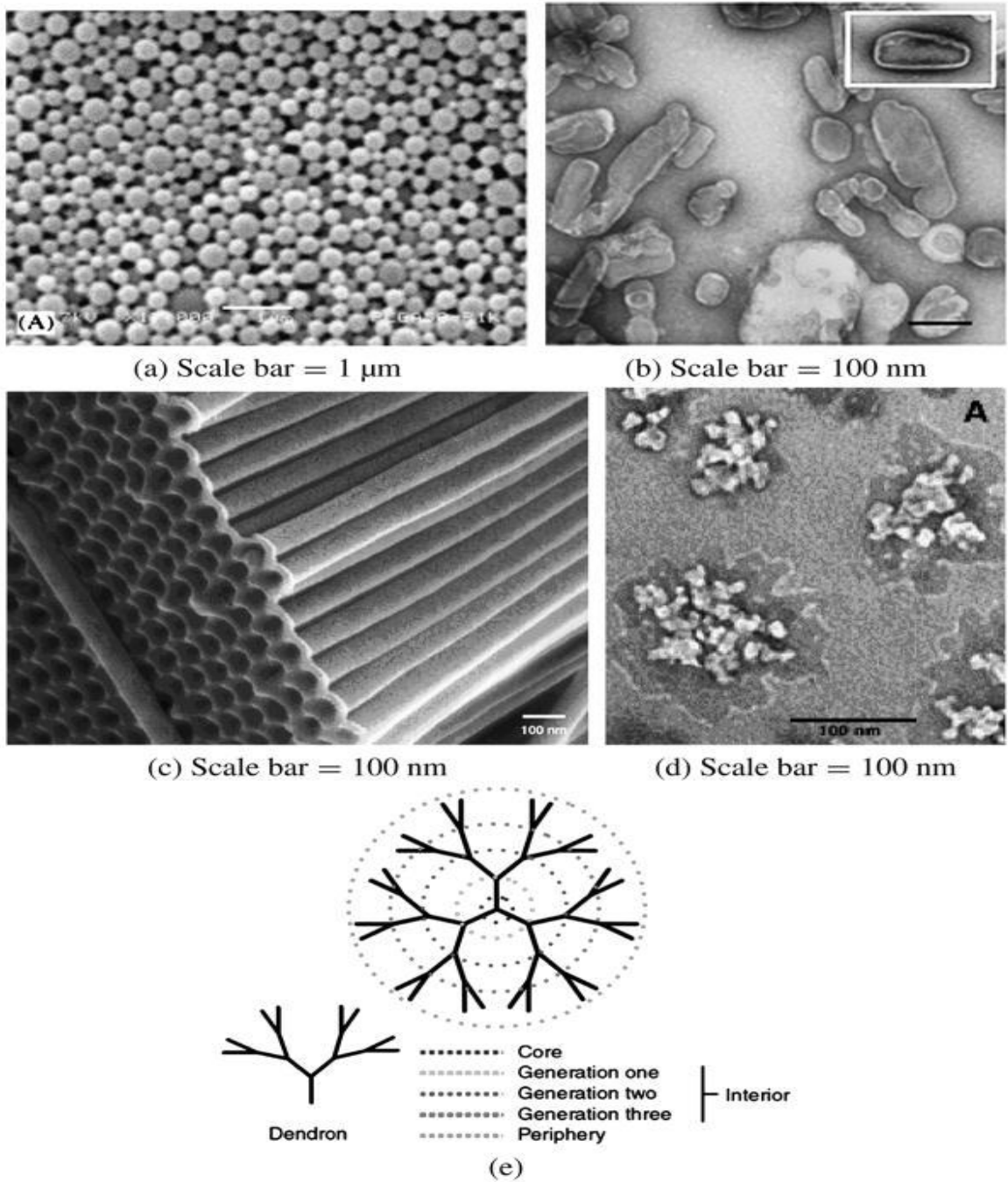

Fig 1.1: (a) Nanoparticles   (b) nanocapsules   (c) nanotubes   (d) nanogels [8]

## 1.1.3 DRUGS HANDLED

### 1.1.3.1 ASPIRIN

Aspirin is also known as acetylsalicylic acid. The IUPAC name is 2-(acetoxy) benzoic acid. Its chemical formula is $C_9H_8O_4$. [9] It is a salicylate drug, mostly used as an analgesic to relieve minor levels of aches and pains. Also used as an antipyretic so as to reduce fever, and as an anti-inflammatory medication. It reduces certain natural substances in the body that cause pain, inflammation and fever. It is sometimes used to treat chest pain (angina) and strokes. It also relieves mild to moderate pain from conditions such as muscle aches, toothaches, common cold, and headaches. Aspirin is also known as a nonsteroidal anti-inflammatory drug (NSAID). People may be directed to take a low dose of aspirin to prevent blood clots. This effect reduces the risk of stroke and heart attack. If there has been a surgery on clogged arteries (such as bypass surgery, carotid endarterectomy, coronary stent), patient may be directed to use aspirin in low doses as a "blood thinner" to prevent blood clots. This is because it has an antiplatelet effect which is inhibiting the production of thromboxane in the blood, which is used to bind platelets together to create a blockage or obstruction over damaged walls of blood vessels for the blood to not ooze out. But this platelet patch can become very large and may block the blood flow, which is why aspirin can be used for a long-term, at low doses, to prevent heart attacks, strokes, and blood clot formation in people at high risk of developing blood clots. [10] Aspirin has certain negative points too. It can cause Reye's syndrome which is a serious and sometimes fatal condition in children. [11] The metabolization is rapid hydrolysation to salicylic acid which is conjugated in the liver to the metabolites. For the treatment of moderate to severe pain it is frequently used along with other opioid analgesic and other non-steroidal-anti-inflammatory drugs. Severe pains can be treated with combination of other different drugs with aspirin. Rheumatic fever and rheumatic arthritis can also be treated to some extent. It can be used in the treatment of pericarditis, coronary artery disease, myocardial infarction and colorectal cancer. Patients diagnosed with colorectal cancer who regularly use aspirin have a lower risk of colorectal cancer compared to patients not using aspirin. Sometimes one medicine can reduce the effect of another medicine - this is called drug interaction. Aspirin interacts with anti-inflammatory painkillers - such as ibuprofen, diclofenac, indomethacin, and naproxen but may also increase the risk of bleeding in stomach if taken along with aspirin. Aspirin also interacts with methotrexate which is used for the treatment of cancer. Aspirin can make it difficult for the body to remove methotrexate, causing highly dangerous levels of methotrexate in body. It reacts with warfarin which is an anticoagulant drug and obstructs the blood clotting. Aspirin if taken with warfarin can also reduce its anticoagulant effects, thus increasing the risk of bleeding.

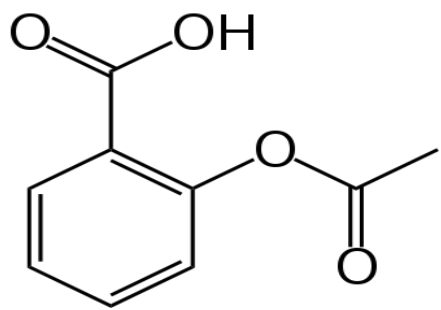

Fig 1.2: Structure of Aspirin [12]

The side effects of aspirin include irritation of the stomach or gut, indigestion, nausea and more severe but less common effects include vomiting, inflammation or bleeding of stomach, bruising and the extremely rare effect due to low dose aspirin may be hemorrhagic stroke. Aspirin is poorly soluble in water and thus can also cause gastrointestinal irritation. About 50–80% of salicylate in the blood binds to albumin, and the rest remains in the active state. The protein binding is depends on concentration. The chemical properties say that it is stable in dry air, but slowly hydrolyses when in contact with moisture to acetic and also salicylic acids. In solution with alkalis, the hydrolysis takes place very fast. Aspirin is used for drug delivery because it is widely used for various conditions. Many researches have been done till date for delivery of drug using aspirin.

Aspirin has never been assigned to pregnant women by the FDA. Although, aspirin is definitely considered in pregnancy category D, if full dose aspirin is administered in the third month. Using nonsteroidal anti-inflammatory drugs in the duration of the third trimester of pregnancy should be avoided thoroughly as it has bad effects on fetal cardiovascular system. Usage in pregnancy is associated with adverse alterations in maternal and fetal hemostasis. High doses are associated with increased perinatal mortality. During the first two months of pregnancy, aspirin should only be administered when clearly needed.

Increased profuse bleeding can take place during delivery if aspirin is used 1 week before and/or while there is labor and delivery. Prolonged labor has been seen in women as aspirin inhibits prostaglandin. [13] [14]

1.1.3.2    STRONTIUM RANELATE

Strontium ranelate comprises of an organic part i.e. ranelic acid and of two atoms of stable non-radioactive strontium. Strontium is a metal naturally found as a non-radioactive element. Almost 99% of the strontium in the human body is concentrated in the bones. Several different forms of strontium are used by scientists for testing if it can be administered orally to treat osteoporosis. Radioactive strontium-89 is administered intravenously for prostate cancer and bone cancer. Strontium chloride hexahydrate is used in toothpaste to reduce pain in sensitive teeth. Strontium chloride is a form found in dietary supplements. These supplements are used for building bones. But there is not much available information about the safety of strontium chloride when taken orally. Strontium ranelate increases bone formation and prevents bone loss in postmenopausal women with osteoporosis. Strontium in dietary has these effects. A radioactive strontium form can kill some cancer cells but this type is not available in dietary supplements. Using strontium for osteoarthritis has developed suggesting it will boost collagen and cartilage formation in joints. There is some interest in study of strontium to prevent tooth decay because there are fewer dental caries in some population groups that drink public water containing high levels of strontium.

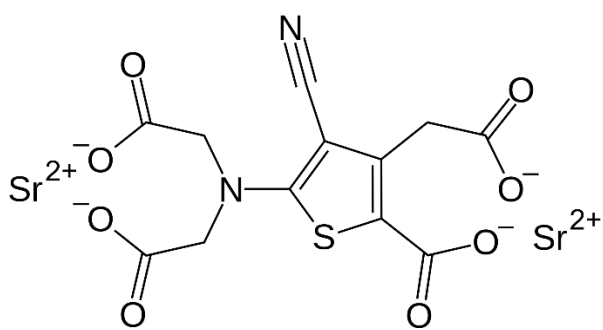

Fig 1.3: Structure of strontium ranelate [15]

It is effective for bone pain related to bone cancer. Strontium-89 chloride when administered intravenously reduces pain in metastatic bone cancer. It is effective for sensitive teeth in a way that using strontium chloride with strontium acetate in toothpaste minimizes pain in sensitive teeth. For osteoporosis evidences show that taking strontium ranelate orally reduces the risk of vertebral and nonvertebral fractures. It increases bone mass in postmenopausal women with osteoporosis. In prostate cancer giving strontium-89 chloride intravenously slows the growth of prostate. In osteoarthritis, taking strontium ranelate orally for 3 years improves backache and prevents spinal osteoarthritis from becoming severe. For dental cavities also it is very useful. More evidence on the basis of research is needed to know the effectiveness of strontium for the aforementioned uses.

Strontium is safe when taken orally in certain amounts found in food. Typical diet should include 0.5 to 1.5 mg/day of strontium. The prescribed form is strontium-89 chloride which is also safe when administered intravenously. Toothpastes that contain strontium have received safety approval from the U.S. Food and Drug Administration (FDA). Taking strontium ranelate orally for 56 months is also safe. It can cause side effects such as diarrhea, stomach pain, and headache. Taking high doses of strontium orally is unsafe. High doses of strontium may damage the bones. There is lack of enough information to know if the form in dietary supplements is safe. Paget's disease which is a bone disease, the bones of people with Paget's disease take up more strontium. In kidney problems strontium is eliminated by the kidneys and so it can build up in people with poor kidney functioning. Use strontium supplements with caution if you have kidney disease. Strontium ranelate should not be used if kidney disease is advanced. In blood clotting disorders, strontium ranelate causes a small increase in risk of blood clots. There is some concern that strontium might be more likely to cause blot clots in people with blood clotting disorders. It's best not to use strontium if you have a clotting disorder. [16]

In vitro, strontium ranelate increases collagen and non-collagenic proteins synthesis by mature osteoblast enriched cells. Strontium ranelate helps bone formation by enhancing pre-osteoblastic cells replication. It stimulates replication of osteoprogenitor cells, collagen and non-collagenic protein synthesis in osteoblasts providing enough evidence to consider SR as a bone forming agent. In the mouse culture system, SR initiates a dose-dependent inhibition of calcium release. These effects show that SR affects bone resorption due to inhibition of osteoclast activity. In normal rats, administration of SR increases improvement in properties of the humerus and lumbar vertebra with an increase in bone dimension, and volume. The combinational intake of SR and calcium reduces the bio-availability of SR. It does not cause any adverse reaction. [17]

### 1.1.4 NATURAL CALCIUM CARBONATE

Humanity has valued sea shells for their beauty as ornaments and utility as tools for thousands of years. But even alchemists never tried transmuting base materials into the very fine interlayering necessary to create a shell's strength, hardness, and toughness.

A seashell or sea shell, also known simply as a shell, is a hard, protective outer layer created by an animal that lives in the sea. The shell is part of the body of the animal. Empty seashells are often found washed up on beaches by beachcombers. The shells are empty because the animal has died and the soft parts have been eaten by another animal or have rotted out.

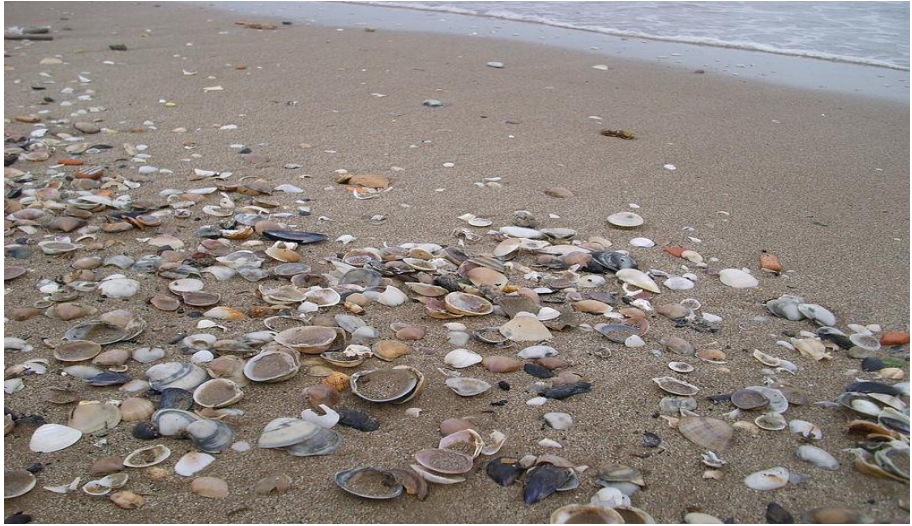

Fig. 1.4: Sea shells

The term seashell usually refers to the exoskeleton of an invertebrate (an animal without a backbone). Most shells that are found on beaches are the shells of marine mollusks, partly because many of these shells endure better than other seashells. [18]

Sea shells are made of calcium carbonate, or $CaCO_3$, which has three common polymorphs, calcite, vaterite and aragonite. Aragonite is slightly more soluble than calcite and also is easier for the animal or plant (there are calcareous algae, as well) to secrete. Fossil shells often have "recrystallized" to calcite, although most sea shells originally are aragonite. The rock Limestone is made of calcium carbonate, in either of the two forms. Lime, however, is CaO, not the material from which shells are made, but is obtained by the process called 'Calcining'.

**Calcite** is a carbonate mineral and the most stable polymorph polymorph of calcium carbonate. The other polymorphs are the minerals aragonite and vaterite. Aragonite will change to calcite at 380–470°C, and vaterite is even less stable.

**Aragonite** is a carbonate mineral, one of the two common, naturally occurring, crystal forms of calcium carbonate (the other form being the mineral calcite). It is formed by biological and physical processes, including precipitation from marine and freshwater environments.

Aragonite's crystal lattice differs from that of calcite, resulting in a different crystal shape, an orthorhombic system with acicular crystals. Repeated twinning results in pseudo-hexagonal forms. Aragonite may be columnar or fibrous, occasionally in branching stalactite forms called flos-ferri ("flowers of iron") from their association with the ores at the Corinthians iron mines.

**Vaterite** is a mineral, a polymorph of calcium carbonate. It was named after the German mineralogist Heinrich Vater. It is also known as mu-calcium carbonate ($\mu$-$CaCO_3$). Vaterite, like aragonite, is a metastable phase of calcium carbonate at ambient conditions at the surface of the earth. As it is less stable than either calcite or aragonite, vaterite has a higher solubility than either of these phases. Therefore, once vaterite is exposed to water, it converts to calcite (at low temperature) or aragonite (at high temperature: ~60 °C). However, vaterite does occur naturally in mineral spirings, organic tissue, gall stones, and urinary calculi. In those circumstances, some impurities (metal ions or organic matter) may stabilize the vaterite and prevent its transformation into calcite or aragonite. Vaterite is usually colourless, its shape is spherical, and its diameter is small, ranging from 0.05 to 5 μm. Vaterite can be produced as the first mineral deposits repairing natural or experimentally induced shell damage in some aragonite-shelled mollusks (e.g. gastropods). Subsequent shell deposition occurs as aragonite.

Calcium is a metallic element, but it is not found in its elemental form in nature due to its reactivity. Calcium tends to react with other elements, producing compounds called ionic salts that consist of positively charged particles of calcium and negatively charged particles of varying identity. In the case of calcium carbonate, the negative particles are carbonate, with the chemical formula $CaCO_3$. [19]

**Applications of calcium carbonate**

Health and dietary applications

Calcium carbonate is widely used medicinally as an inexpensive dietary calcium supplement or gastric antacid. It may be used as a phosphate binder for the treatment of hyperphosphatemia (primarily in patients with chronic renal failure). It is also used in the pharmaceutical industry as an inert filler for tablets and other pharmaceuticals.

Calcium carbonate is known among IBS sufferers to help reduce diarrhoea. Some individuals report being symptom-free since starting supplementation. The process in which calcium carbonate reduces diarrhoea is by binding water in the bowel, which creates a stool that is firmer and better formed. Calcium carbonate supplements are often combined with magnesium in various proportions. This should be taken into account as magnesium is known to cause diarrhoea.

Calcium carbonate is used in the production of toothpaste and has seen resurgence as a food preservative and colour retainer, when used in or with products such as organic apples or food.

Excess calcium from supplements, fortified food and high-calcium diets, can cause the milk-alkali syndrome, which has serious toxicity and can be fatal. In 1915, Bertram Sippy introduced the "Sippy regimen" of hourly ingestion of milk and cream, and the gradual addition of eggs and cooked cereal, for 10 days, combined with alkaline powders, which provided symptomatic relief for peptic ulcer disease. Over the next several decades, the Sippy regimen resulted in renal failure, alkalosis, and hypercalcemia, mostly in men with peptic ulcer disease. These adverse effects were reversed when the regimen stopped, but it was fatal in some patients with protracted vomiting. Milk alkali syndrome declined in men after effective treatments for peptic ulcer disease arose. During the past 15 years, it has been reported in women taking calcium supplements above the recommended range of 1.2 to 1.5 gm daily, for prevention and treatment of osteoporosis, and is exacerbated by dehydration. Calcium has been added to over-the-counter products, which contributes to inadvertent excessive intake. Excessive calcium intake can lead to hypercalcemia, complications of which include vomiting, abdominal pain and altered mental status. [18] [19]

Environmental applications

In 1989, a researcher, Ken Simmons, introduced $CaCO_3$ into the Whetstone Brook in Massachusetts. His hope was that the calcium carbonate would counter the acid in the stream from acid rain and save the trout that had ceased to spawn. Although his experiment was a success, it did increase the amount of aluminium ions in the area of the brook that was not treated with the limestone. This shows that $CaCO_3$ can be added to neutralize the effects of acid rain in river ecosystems. Currently calcium carbonate is used to neutralize acidic conditions in both soil and water. Since the 1970s, such *liming* has been practiced on a large scale in Sweden to mitigate acidification and several thousand lakes and streams are limed repeatedly. [19]

Industrial applications

The main use of calcium carbonate is in the construction industry, either as a building material or limestone aggregate for road building or as an ingredient of cement or as the starting material for the preparation of builder's lime by burning in a kiln. However, due to weathering mainly caused by acid rain, calcium carbonate (in limestone form) is no longer used for building purposes on its own, and only as a raw/primary substance for building materials. Calcium carbonate is also used in the purification of iron from iron ore in a blast furnace. The carbonate is calcined *in situ* to give calcium oxide, which forms a slag with various impurities present, and separates from the purified iron.

In the oil industry, calcium carbonate is added to drilling fluids as a formation-bridging and filter cake-sealing agent; it is also a weighting material which increases the density of drilling fluids to control the downhill pressure. Calcium carbonate is added to swimming pools, as a pH corrector for maintaining alkalinity and offsetting the acidic properties of the disinfectant agent.

Calcium carbonate has traditionally been a major component of blackboard chalk. However, modern manufactured chalk is mostly gypsum, hydrated calcium sulphate $CaSO_4 \cdot 2H_2O$. Calcium carbonate is a main source for growing Secrete, or Biorock. Precipitated calcium carbonate (PCC), pre-dispersed in slurry form, is a common filler material for latex gloves with the aim of achieving maximum saving in material and production costs.

Fine ground calcium carbonate (GCC) is an essential ingredient in the micro porous film used in babies' diapers and some building films as the pores are nucleated around the calcium carbonate particles during the manufacture of the film by biaxial stretching. GCC or PCC is used as filler in paper because they are cheaper than wood fibre. Printing and writing paper can contain 10–20% calcium carbonate. In North America, calcium carbonate has begun to replace kaolin in the production of glossy paper. Europe has been practicing this as alkaline papermaking or acid-free papermaking for some decades. PCC has a very fine and controlled particle size, on the order of 2 micrometres in diameter, useful in coatings for paper.

Calcium carbonate is widely used as an extender in paints, in particular matte emulsion paint where typically 30% by weight of the paint is either chalk or marble. It is also popular filler in plastics. Some typical examples include around 15 to 20% loading of chalk in unplasticized polyvinyl chloride (uPVC) drain pipe, 5 to 15% loading of stearate coated chalk or marble in uPVC window profile. PVC cables can use calcium carbonate at loadings of up to 70 phr (parts per hundred parts of resin) to improve mechanical properties (tensile strength and elongation) and electrical properties (volume resistivity). Polypropylene compounds are often filled with calcium carbonate to increase rigidity, a requirement that becomes important at high use temperatures. Here the percentage is often 20–40%. [18] [19] [20]

# CHAPTER 2

## 2.1 EXPERIMENTAL: MATERIALS AND METHODS

**2.1.1 MATERIALS**

**2.1.1.1 Chemicals and Reagents**

- Calcium carbonate powder ($CaCO_3$) was made from sea shells that were collected from Shivaji Park seashore, Dadar, Mumbai. They were thoroughly sonicated and cleaned.

- Aspirin and Strontium ranelate were obtained as a gift from Glenmark Pharmaceuticals LTD. R & D Centre, Malegaon, Sinnar – batch no. of aspirin - 081201535, batch no. of strontium ranelate – 081202368

- Dialysis membrane from HiMedia Laboratories Pvt. Ltd – batch no.- 0000190882

- Water used was of HPLC grade and all other chemicals of analytical grade were used.

**2.1.1.2 Instruments**

- Sonicator - REMI RM12C

- Weighing machine-Sartorius Weighing Technology GmbH, Goettingen, Germany - batch no.CPA225D

- Field Emission Gun-Scanning Electron Microscopes (FE-SEM) - Model: JSM-7600F

- X-Ray Diffraction (XRD) - Model: P analytical MRD System for Bulk Texture and Residual Stress Measurement with specialty as X-Ray lens

- High Resolution Liquid Chromatograph Mass Spectrometer (HR-LCMS) - Model: 1290 Infinity UHPLC System 1260 infinity Nano HPLC with Chipcube, 6550 iFunnel Q-TOF. By Agilent Technologies, USA

- Grinder

- Sieve–REMI (43–150 mirons)

- Magnetic stirrer - REMI (1MLH)

## 2.1.2 PREPARATION OF CaCO$_3$ NANOSPHERES

The sea shells were grinded in a heavy grinder to obtain a very fine powder of the same which comprises of nanoparticles of a humungous size range. This is why the powder was then sieved in a mechanical sieve that gave sizes of nanoparticles ranging from 43 – 150 microns. Nanoparticles of the size below 50 were taken. This powder was stored in a contamination free environment. Size, shape, and composition of natural calcium carbonate particles were of interest.

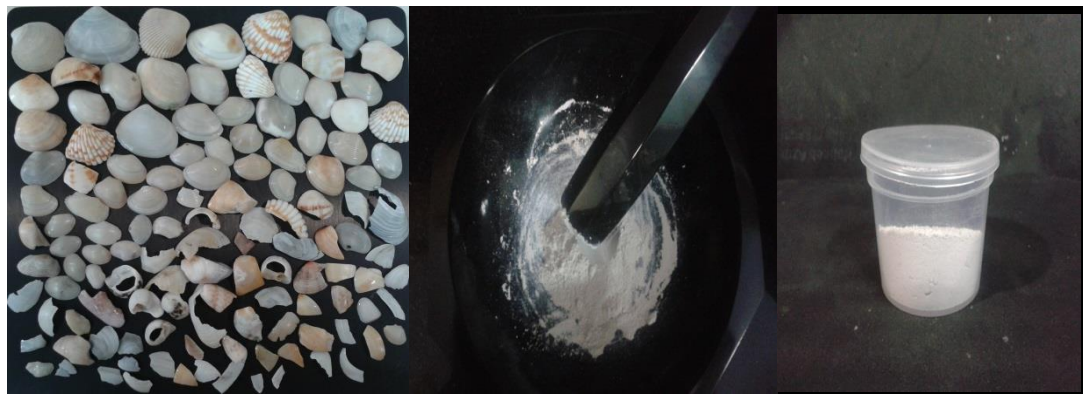

Fig 2.1: Sea shells grinded into powder and collected

## 2.1.3 PREPARATION OF PBS BUFFER

For phosphate buffer saline (PBS) the protocol describes a generalized method for the preparation, sterilization, and storage of phosphate buffered saline. The pH is generally adjusted to 7.4. There are many variant recipes for PBS that include different components or are adjusted to a different pH. Following are the steps:

1. Absolutely clean flasks and beakers washed with dilute HCl first and DI water later obtained.

2. Measure a volume of 800 ml of distilled water of HPLC grade and transfer to a 5000 ml beaker.

3. Add a magnetic stir bar to the beaker and place the flask on a magnetic stir plate. Adjust the speed of the magnetic stir bar so that oxygen is not introduced into the solution while it is rapidly mixed.

4. Transfer to the beaker:

     8 g of NaCl, 0.2 g KCl

     1.44 g of $Na_2HPO_4$

     0.25 g of $KH_2PO_4$

5. Allow the solutes to dissolve for 3 to 5 min.

6. Ensure that there are no remaining particles of undissolved salts in the solution before adjusting the pH. If particles are present, continue stirring vigorously.

7. Reduce the speed of the magnetic stir bar so that the solution is gently mixing.

8. Ensure that the pH meter has been properly calibrated and rinse the pH probe with HPLC grade distilled water. Remove the excess water from the probe tip (without touching the probe tip) with a clean paper towel. Place the pH probe into the solution.

9. Slowly add 1 M HCl dropwise with a pipette and allow the HCl to fully dissolve into the solution.

10. Measure the pH with the pH meter.

11. Repeat Steps #8 through #10 until the pH of the solution is 7.4.

12. Pour the solution into a fresh graduated cylinder and adjust the final volume to 1 liter with distilled water.

13. Store the PBS solution at room temperature. The PBS Solution is sterile; when using the PBS Solution, ensure that sterile techniques are employed.

## 2.1.4 DRUG LOADING

For loading of aspirin firstly a 50 ml stock solution was prepared in ethanol. This is because aspirin dissolves only in ethanol.

### 2.1.4.1 ASPIRIN

**Preparation of Aspirin stock solution**:

500 mg of aspirin was added to 50 ml ethanol.

**Preparation of Aspirin solutions for loading of CaCO$_3$:**

Dilutions of various concentrations ranging from 500 µg/ml – 10,000 µg/ml. These solutions of different concentrations were made by adding water to the aspirin stock solution. Below is the table giving proper values for the same:

| Concentrations (µg/ml) | Aspirin stock solution (ml) | Water (ml) |
|---|---|---|
| 500 | 2 | 38 |
| 1000 | 4 | 36 |
| 2000 | 8 | 32 |
| 5000 | 20 | 20 |
| 10000 | 5 | - |

Table 2.1: Diluted aspirin solutions for loading of CaCO$_3$.

**Drug loading of aspirin:**

The sieved CaCO$_3$ powder of the size below 50 microns was taken, 100 mg of which was taken in five different 10ml beakers and add the diluted aspirin solution to each of the beakers.

The amount of CaCO$_3$ powder needs to be in a specific ratio with the concentration or volume of the drug taken. This is for the drug to get entrapped within the nanoparticles it is important to take same, double or multiple times more of the drug than the amount of CaCO$_3$ powder taken.

The solutions were mixed by shaking first then sonicated for about 15-30 mins. These solutions were then kept for about 48 hours for the drug to get settled and entrapped in the CaCO$_3$ powder. The water content evaporates slowly leaving the drug, NaCl and CaCO$_3$ powder behind. It is important to keep in mind that the solutions must not be disturbed or receive any sudden jerk, which may sometimes cause the drug to come out of CaCO$_3$.

| CaCO$_3$ powder (mg) | Diluted aspirin solutions (ml) | Aspirin quantity (mg) | % of Aspirin |
|---|---|---|---|
| 100 | 5 | 2.5 | 2.5 |
| 100 | 5 | 5 | 5 |
| 100 | 5 | 10 | 10 |
| 100 | 5 | 25 | 25 |
| 100 | 5 | 50 | 50 |

Table 2.2: Drug loading of aspirin

After the solution is evaporated leaving only dry layer of powder at the bottom, scrape it and collect in small 1 ml Eppendorf tubes. Close them tightly to avoid moisture content from increasing.

2.1.4.2    STRONTIUM RANELATE

**Preparation of SR stock solution**:

Strontium ranelate (SR) dissolves only in NaCl solution. The standard NaCl solution preparation is 0.9 gm in 100 ml distilled water. Here, take 0.36 gm of NaCl powder in 40 ml distilled water. Then, add 10 ml of distilled water to this 40 ml NaCl solution. Dissolve 500 mg SR in this 50 ml diluted NaCl solution by stirring for about 30 mins. The solution should look clear in appearance.

**Preparation of SR solutions for loading of $CaCO_3$:**

Dilutions of various concentrations ranging from 500 µg/ml – 10,000 µg/ml. These solutions of different concentrations were made by adding water to the SR stock solution. Below is the table giving proper values for the same:

| Concentrations (µg/ml) | SR stock solution (ml) | Water (ml) |
|---|---|---|
| 500 | 2 | 38 |
| 1000 | 4 | 36 |
| 2000 | 8 | 32 |
| 5000 | 20 | 20 |
| 10000 | 5 | - |

Table 2.3: Diluted SR solutions for loading of $CaCO_3$.

**Drug loading of SR:**

The sieved $CaCO_3$ powder of the size below 50 microns was taken, 100 mg of which was taken in five different 10 ml beakers and add the diluted aspirin solution to each of the beakers.

The amount of $CaCO_3$ powder needs to be in a specific ratio with the concentration or volume of the drug taken. This is for the drug to get entrapped within the nanoparticles it is important to take same, double or multiple times more of the drug than the amount of $CaCO_3$ powder taken.

The solutions were mixed by shaking first then sonicated for about 15-30 mins. These solutions were then kept for about 48 hours for the drug to get settled and entrapped in the $CaCO_3$ powder. The water content evaporates slowly leaving the drug, NaCl and $CaCO_3$ powder behind. It is important to keep in mind that the solutions must not be disturbed or receive any sudden jerk, which may sometimes cause the drug to come out of $CaCO_3$.

| $CaCO_3$ powder (mg) | Diluted SR solutions (ml) | SR quantity (mg) | % of SR |
|---|---|---|---|
| 100 | 5 | 2.5 | 2.5 |
| 100 | 5 | 5 | 5 |
| 100 | 5 | 10 | 10 |
| 100 | 5 | 25 | 25 |
| 100 | 5 | 50 | 50 |

Table 2.4: Drug loading of SR

After the solution is evaporated leaving only dry layer of powder at the bottom, scrape it and collect in small 1 ml Eppendorf tubes. Close them tightly to avoid moisture content from increasing.

## 2.1.5 CHARACTERIZATION AND QUANTIFICATION

### 2.1.5.1 XRD

About 2-3 mg of $CaCO_3$ powder, drug loaded $CaCO_3$ powder of all the concentrations were taken.

### 2.1.5.2 FE- SEM

**Sample preparation:**

The samples of $CaCO_3$ powder and drug loaded $CaCO_3$ powder of all the concentrations were taken. Each powder was taken in very minute quantity, enough to be seen on a piece of butter paper. Another small piece of butter paper was taken and rubbed in circular motion over the powder to get rid of any agglomeration if present. This causes a thin film formation of the powder. The metal stubs with a carbon tape on it having the sticky side up, is touched over this film causing the particles to stick to the carbon tape.

These stubs with samples on them were then kept in the platinum coating machine. The machine uses vacuum environment to bombard the sample particles with platinum particles. This after about 5-10 mins, there forms a coating of platinum metal over the sample. The main use of this is for better conductivity for electrons to pass while viewing the images. The better the conductivity, better are the images.

### 2.1.5.3 HR-LCMS

Liquid chromatography / Mass Spectroscopy (LC / MS) is a technique which combines high performance liquid chromatography HPLC, a powerful analytical separation technique with mass spectroscopy, a powerful analysis & detection technique. There are two common atmospheric pressure ionization (API) LC/MS process: Electrospray Ionization (ESI) & Atmospheric Pressure Chemical Ionization (APCI). Both are soft ionization technique. The technique combines highly efficient electrospray ion generation and focusing of Agilent Jet Stream technology with a hexabore capillary sampling array and dual-stage ion funnel for increased ion sampling and transmission. Nano HPLC combined with mass spectrometer can analyzed small molecule as well large molecules like proteins UHPLC separations can be detected by PDA & Mass spectrometer as different detectors.

The samples mentioned in Buffer sample collection in section 3.1.6.1 and 3.1.6.2 were collected in the small vials that have a rubbery entrance in the cap for sample injection. The solutions were filtered before transferring them in the vials.

### 2.1.5.4   UV-Visible Spectroscopy

This technique was specially used for Strontium ranelate as it has proven to be of high concentration for HR-LCMS. Less diluted samples are required for UV compared to the LCMS. Here mainly this technique is used for checking drug release.

Different concentrations like 5 µg/ml, 25 µg/ml, 50 µg/ml, 100 µg/ml, 200 µg/ml of the drug (SR) were prepared from the same stock solution. These were subjected to optical density measurements at 323 nm by the UV-Vis spectrophotometer and a calibration curve was plotted.

Then the supernatant of the loaded mixture of SR and $CaCO_3$ was filtered and diluted with distilled water 5 times its volume and the filtrates were subjected to optical density tests at the same wavelength. With the help of the resultant peaks, slope and intercept, the percent efficiency was calculated.

## 2.1.6 DRUG RELEASE

Drugs that are once entrapped take different amount of time to get release with respect to different concentrations. Dialysis technique was used for releasing the drug. Consideration of the interactions of drugs and dialysis must include an understanding of the mechanism of transport during dialysis, i.e., diffusion.

### 2.1.6.1 ASPIRIN

The drug was taken in the quantities mentioned in Table 2.5, inside the five separate dialysis bags which were tied up on both the ends and hung in five separate 100 ml beakers. Take 50 ml of PBS buffer in each of the beakers. The beakers contained small stirrer bars and kept on the magnetic stirrers.

The aspirin now starts getting released through the permeable membrane into the buffer due to diffusion. The buffer samples were collected after every one hour for 5 hrs straight. The last buffer sample was collected after 19 hrs. Thus there were 5 samples of 5 different concentrations of 5 hrs and also 5 samples of the last hour.

**Buffer sample collection:**

For every hour, 0.5 ml of sample was directly collected from the concentrations 500 µg/ml and 1000 µg/ml. However, every hour collection for the concentrations 2000 µg/ml, 5000 µg/ml and 10000 µg/ml, only 0.1 ml of the buffer sample was taken and to this, 0.4 ml of fresh PBS buffer was added. This was done to dilute the samples and keep the levels of NaCl low, so that it does not interfere much during LCMS.

| Sample concentration (μg/ml) | Quantity of loaded CaCO$_3$ (mg) | Total initial aspirin quantity in loaded CaCO$_3$ (mg) | PBS buffer in beaker (ml) |
|---|---|---|---|
| 500 | 40 | 1 | 50 |
| 1000 | 20 | 1 | 50 |
| 2000 | 10 | 2 | 50 |
| 5000 | 10 | 5 | 50 |
| 10000 | 10 | 10 | 50 |

Table 2.5: Aspirin dialysis

**Preparation of standards for aspirin:**

| Concentration (μg) | Aspirin (μg) | PBS buffer (μl) |
|---|---|---|
| 5 | 5 | 500 |
| 10 | 10 | 500 |
| 15 | 15 | 500 |
| 20 | 20 | 500 |

Table 2.6: Standards for aspirin

2.1.6.2    STRONTIUM RANELATE

The drug was taken in the quantities mentioned in Table 2.7, inside the five separate dialysis bags which were tied up on both the ends and hung in five separate 100 ml beakers. Take 50 ml of PBS buffer in each of the beakers. The beakers contained small stirrer bars and kept on the magnetic stirrers. The aspirin now starts getting released through the permeable membrane into the buffer due to diffusion. The buffer samples were collected after every one hour for 5 hrs straight. The last buffer sample was collected after 19 hrs. Thus there were 5 samples of 5 different concentrations of 5 hrs and also 5 samples of the last hour.

**Buffer sample collection:**

For every hour, 0.5 ml of sample was directly collected from the concentrations 500 µg/ml and 1000 µg/ml. However, every hour collection for the concentrations 2000 µg/ml, 5000 µg/ml and 10000 µg/ml, only 0.1 ml of the buffer sample was taken and to this, 0.4 ml of fresh PBS buffer was added. This was done to dilute the samples and keep the levels of NaCl low, so that it does not interfere much during LCMS.

| Sample concentration (µg/ml) | Quantity of loaded $CaCO_3$ (mg) | Total initial SR quantity in loaded $CaCO_3$ (mg) | PBS buffer in beaker (ml) |
|---|---|---|---|
| 500 | 40 | 1 | 50 |
| 1000 | 20 | 1 | 50 |
| 2000 | 10 | 2 | 50 |
| 5000 | 10 | 5 | 50 |
| 10000 | 10 | 10 | 50 |

Table 2.7: SR dialysis

**Preparation of standards for SR:**

| Concentration (µg) | SR (µg) | NaCl solution (µl) | PBS buffer (µl) |
|---|---|---|---|
| 5 | 5 | 1 | 499 |
| 10 | 10 | 1 | 499 |
| 15 | 15 | 1 | 499 |
| 20 | 20 | 1 | 499 |

Table 2.8: Standards for SR

# CHAPTER 3

## 3.1 RESULTS AND DISCUSSION

## 3.1.1 CHARACTERIZATION AND QUANTIFICATION

### 3.1.1.1 X-Ray Diffraction (XRD)

Calcium carbonate nanoparticles are abundant inorganic biomaterials with different morphological structures that have attracted the interest of researchers in different fields. The Calcium carbonate occurring in nature contains Aragonite. Aragonite is soluble and thus easier for the animals or plant (there are calcareous algae, as well) to secrete. Fossil shells often have "recrystallized" to calcite, although most sea shells originally are aragonite. X-ray diffraction is a sensitive instrument used for the identification of crystalline phases of inorganic compounds. X-ray powder diffraction analysis was performed.

X-ray diffraction analysis is the method by which multiple beams of X-ray create a three dimensional picture of the density of electrons of any crystalline structure. The purpose is to identify with a high degree of certainty, the composition of the molecules, on an atomic scale. This makes it the most reliable method to determine the purity of calcium carbonate.

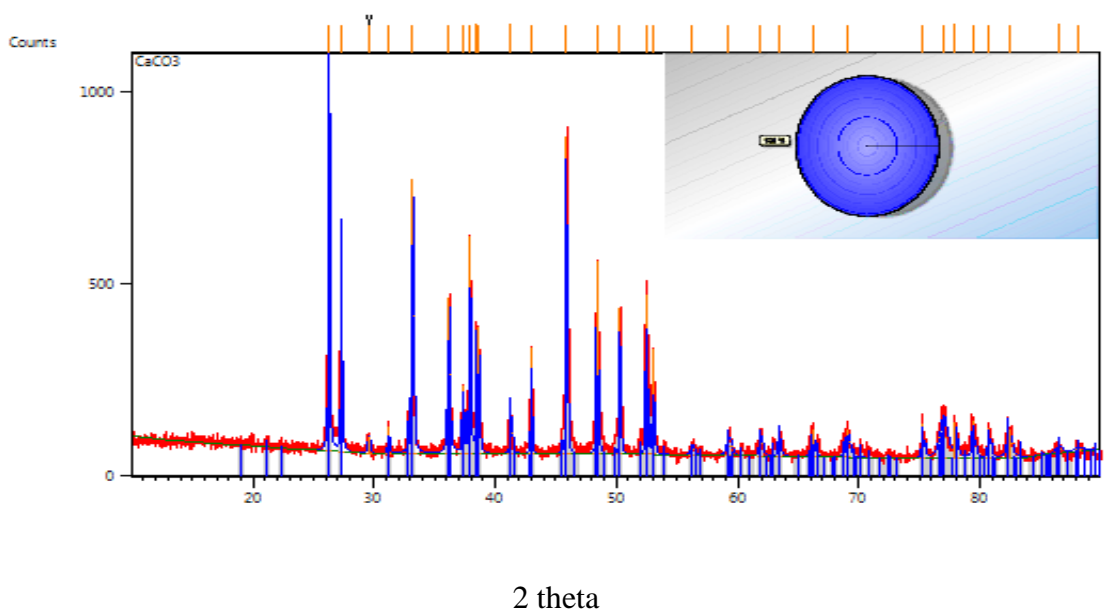

2 theta

Fig 3.1: Shows the XRD pattern of natural calcium carbonate obtained from the sea shells.

The XRD pattern of calcium carbonate obtained from the sea shells is shown in fig 3.1, with blue colour. The blue pie diagram shows that it is 100% calcium carbonate.

**3.1.1.2   Field Emission Scanning Electron Microscopy (FE-SEM) and (EDX)**

In FE-SEM an electron beam is scanned across a sample's surface. When the electrons strike the sample, a variety of signals are generated, and it is the detection of specific signals which produces an image or a sample's elemental composition. The three signals which provide the greatest amount of information in FE-SEM are the secondary electrons, backscattered electrons, and X-rays. Secondary electrons are emitted from the atoms occupying the top surface and produce a readily interpretable image of the surface. The contrast in the image is determined by the sample morphology. A high resolution image can be obtained because of the small diameter of the primary electron beam.

Backscattered electrons are primary beam electrons which are 'reflected' from atoms in the solid. The contrast in the image produced is determined by the atomic number of the elements in the sample. The image will therefore show the distribution of different chemical phases in the sample. Because these electrons are emitted from a depth in the sample, the resolution in the image is not as good as for secondary electrons. Interaction of the primary beam with atoms in the sample causes shell transitions which result in the emission of an X-ray. The emitted X-ray has an energy characteristic of the parent element. Detection and measurement of the energy permits elemental analysis (Energy Dispersive X-ray Spectroscopy or EDS). EDS can provide rapid qualitative, or with adequate standards, quantitative analysis of elemental composition with a sampling depth of 1-2 microns. X-rays may also be used to form maps or line profiles, showing the elemental distribution in a sample surface. [21]

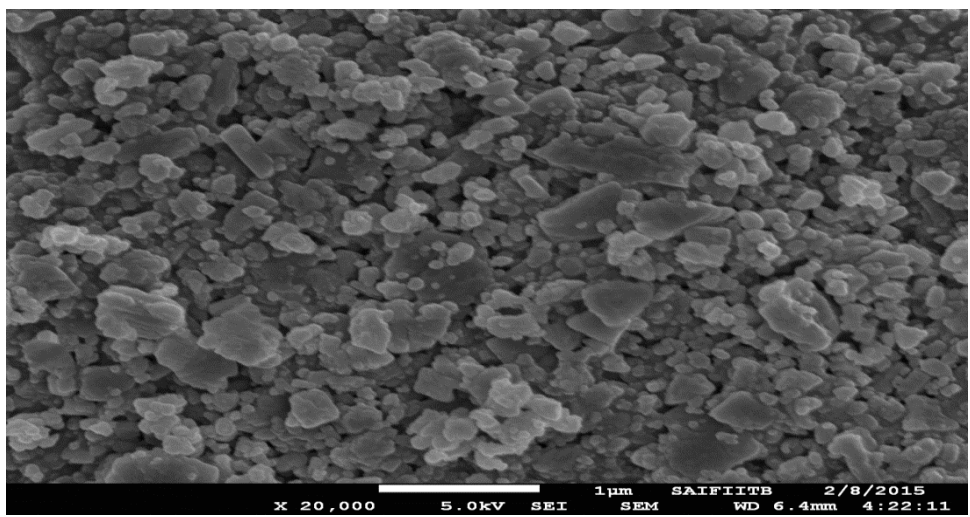

Fig 3.2: FE-SEM image of Calcium carbonate

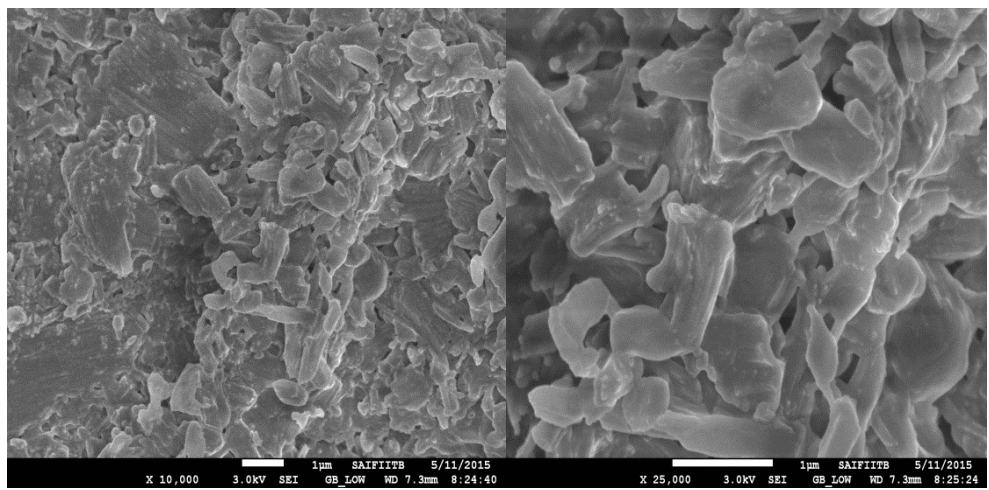

Fig 3.3: FE-SEM images of Aspirin

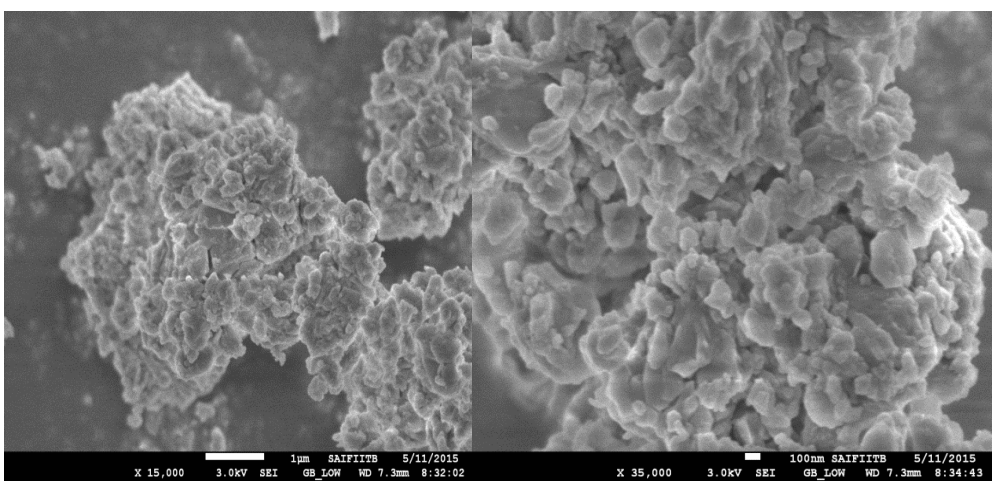

Fig 3.4: FE-SEM images of Strontium ranelate

**Energy dispersive X-Ray (EDX) composition analysis**

Energy Dispersive X-Ray Analysis (EDX), referred to as EDS or EDAX, is an x-ray technique used to identify the elemental composition of materials.

EDX systems are attachments to Electron Microscopy instruments like FE-SEM or Transmission Electron Microscopy (TEM), where the imaging capability of the microscope identifies the specimen of interest. The data generated by EDX analysis consist of spectra showing peaks corresponding to the elements making up the true composition of the sample being analyzed. Elemental mapping of a sample and image analysis are also possible.

In a multi-technique approach EDX becomes very powerful, particularly in contamination analysis and industrial forensic science investigations. The technique can be qualitative, semi-quantitative, and quantitative; also providing spatial distribution of elements through mapping. The EDX technique is non-destructive and specimens of interest can be examined in situ with little or no sample preparation. [20]

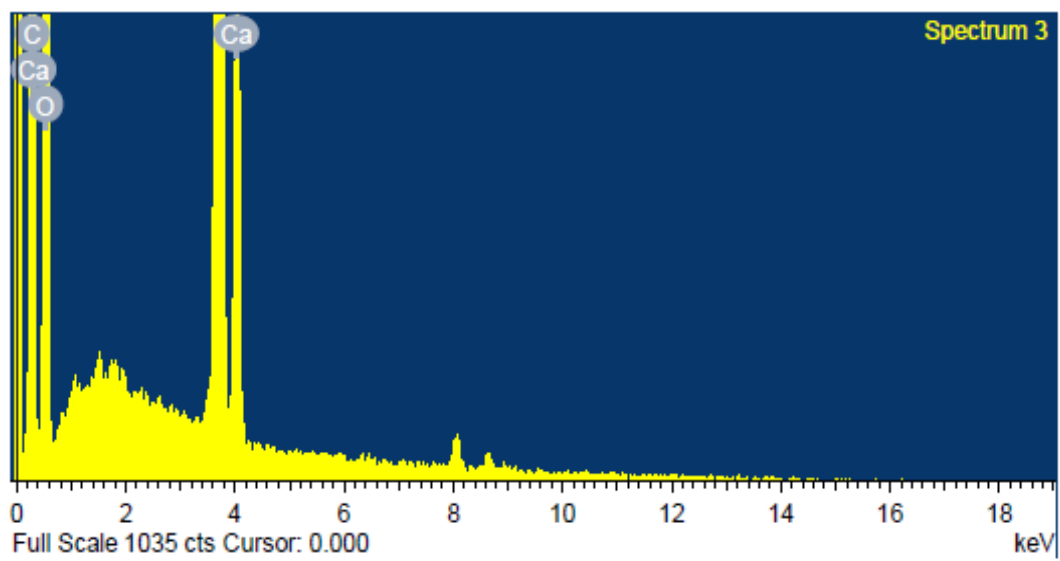

Fig 3.5: Shows the weight % of chemical composition of $CaCO_3$

### 3.1.2 DRUG LOADING STUDIES

In recent years, the development of analytical methods for simultaneous determinations of drugs has gained considerable attention due to their importance in quality control testing of drugs and their products and in ordinary laboratories because of their wide availability and suitability. Since no method is reported in literature for the simultaneous quantification of aspirin and strontium in tablets for the routine quality control assay in ordinary laboratories, the development of such method could be appreciable.

### 3.1.2.1 ASPIRIN

The drug content was evaluated for all the concentrations and it was observed that the nanoparticles of only $CaCO_3$ that showed inter-particle spaces and crevices were now appeared filled up. The drug thus is entrapped within or between the nanoparticles. These results were mainly attributed to aspirin powders increasing the charge density. Various solvent systems like distilled water, ethanol and mixture of ethanol: water were tried and the ethanol: water proved to give the best loading of drug. Maximum loading obtained in this work is 50%. The water is used because it provides some hold to the ethanol moiety for better loading of drug. It makes the process of evaporation slower to a good extent causing lesser harm to the nanoparticle's structures. Below are the FE-SEM images that show increase in the drug loading with an increase in the drug concentration while keeping the quantity of calcium carbonate powder same for all formulations.

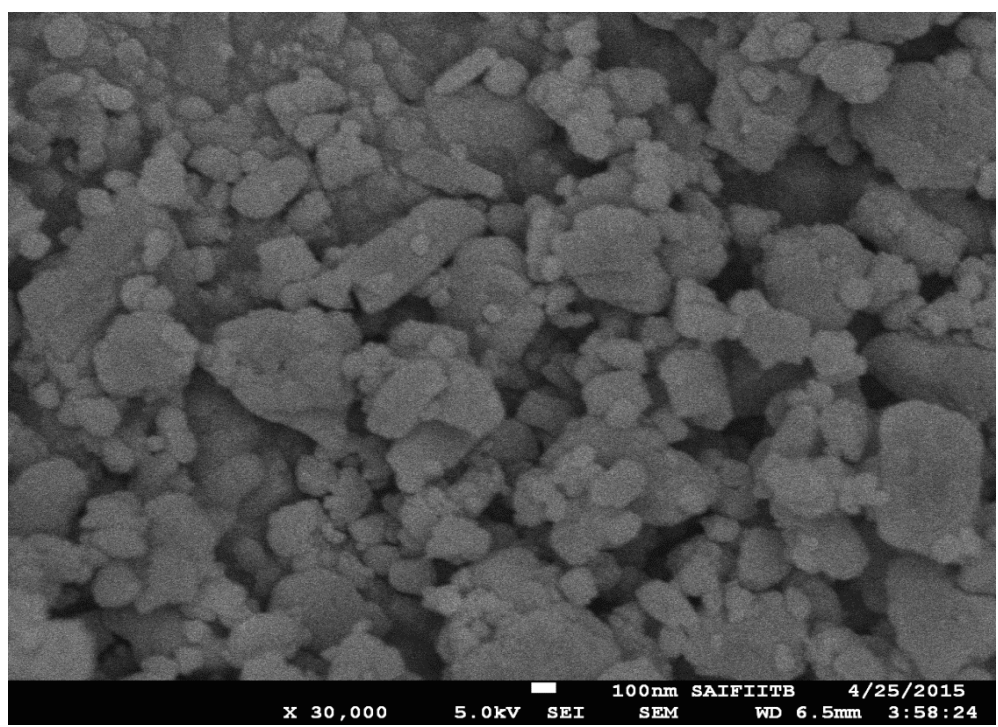

Fig 3.6: Aspirin loading of ratio 1:16 as per Table 4.2

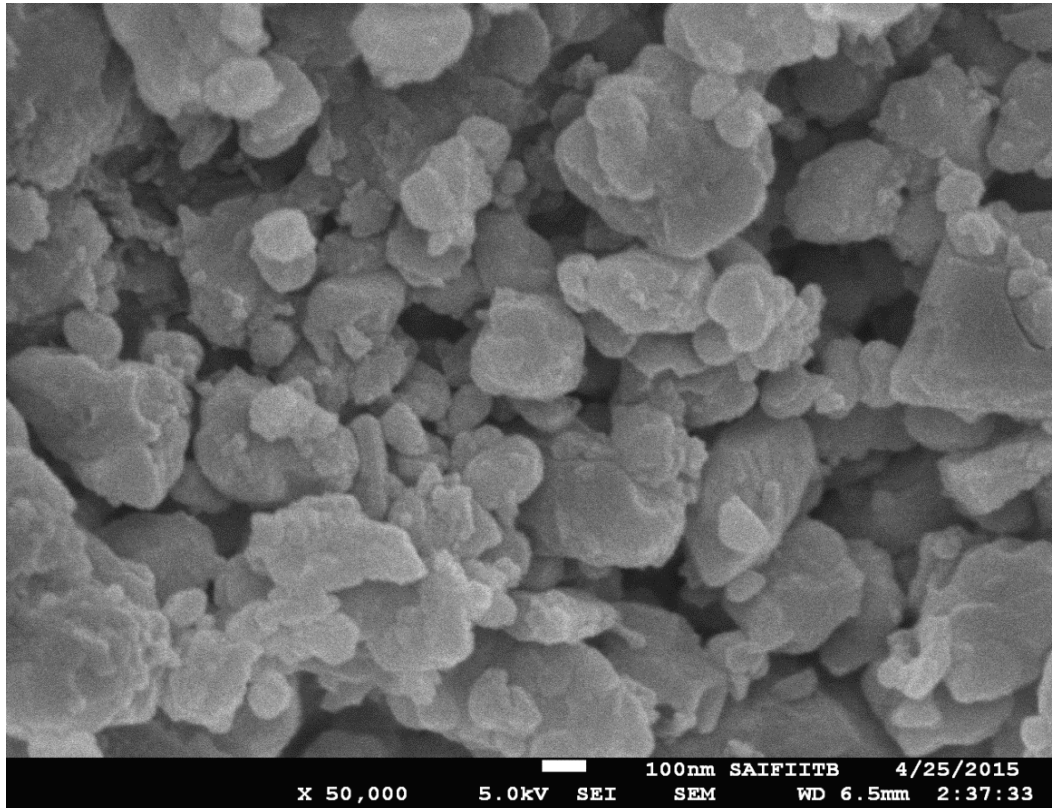

Fig 3.7: Aspirin loading of ratio 1:4 as per Table 4.2

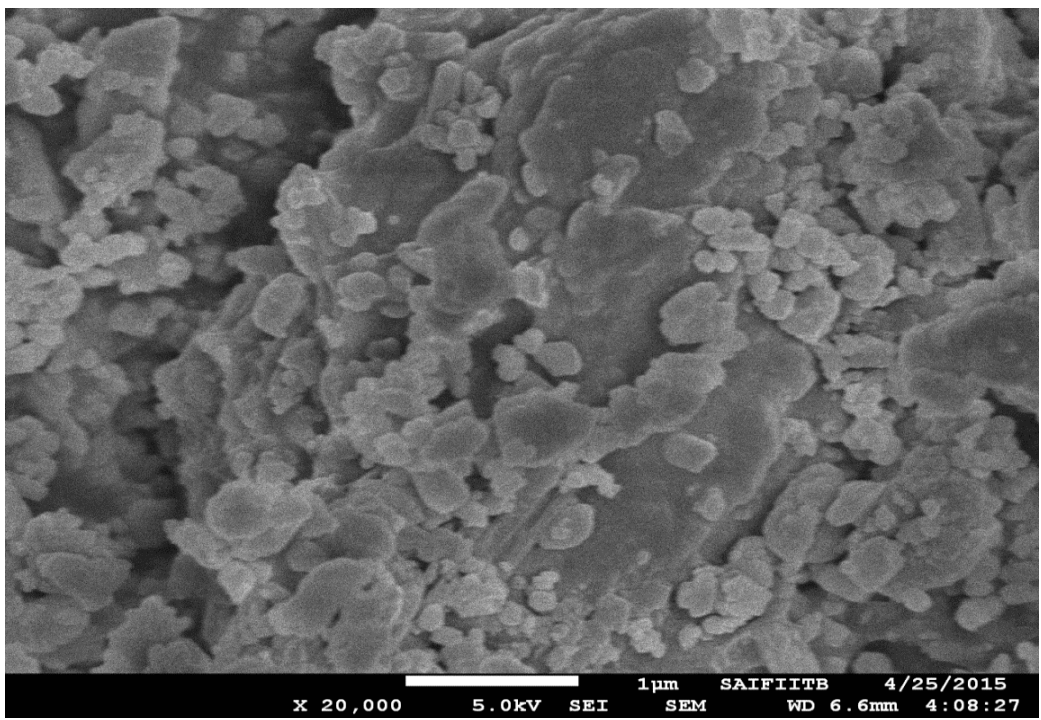

Fig 3.8: Aspirin loading of ratio 1:1 as per Table 4.2

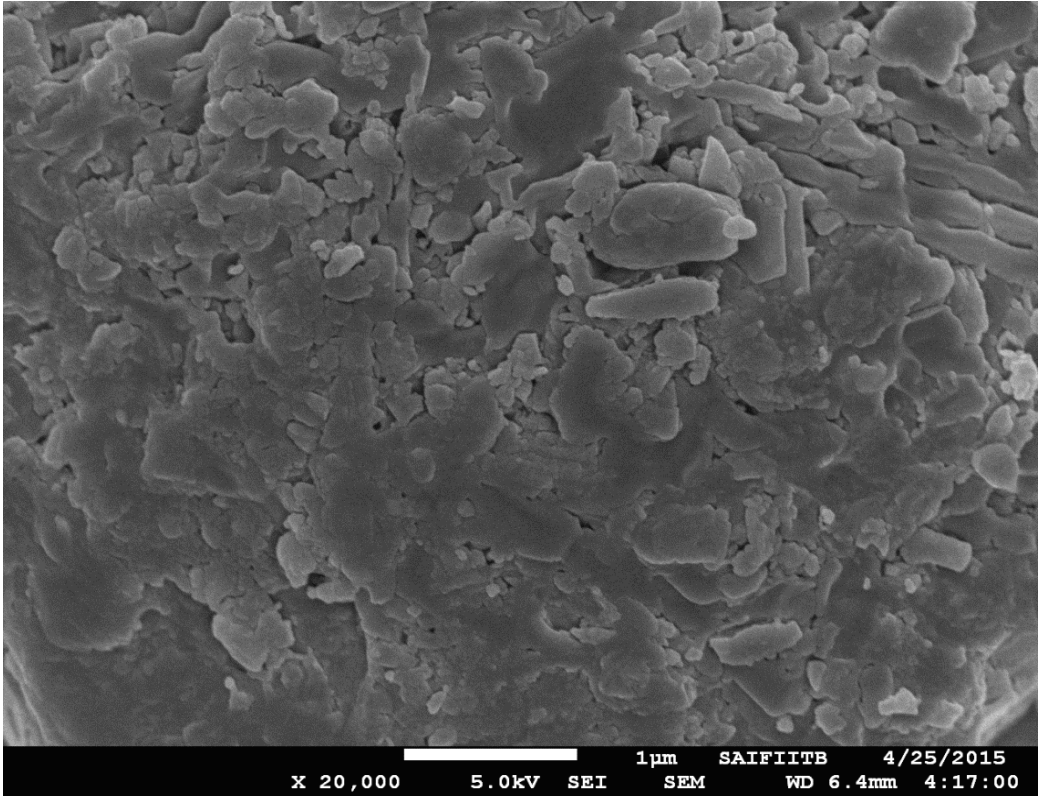

Fig 3.9: Aspirin loading of ratio 5:2 as per Table 4.2

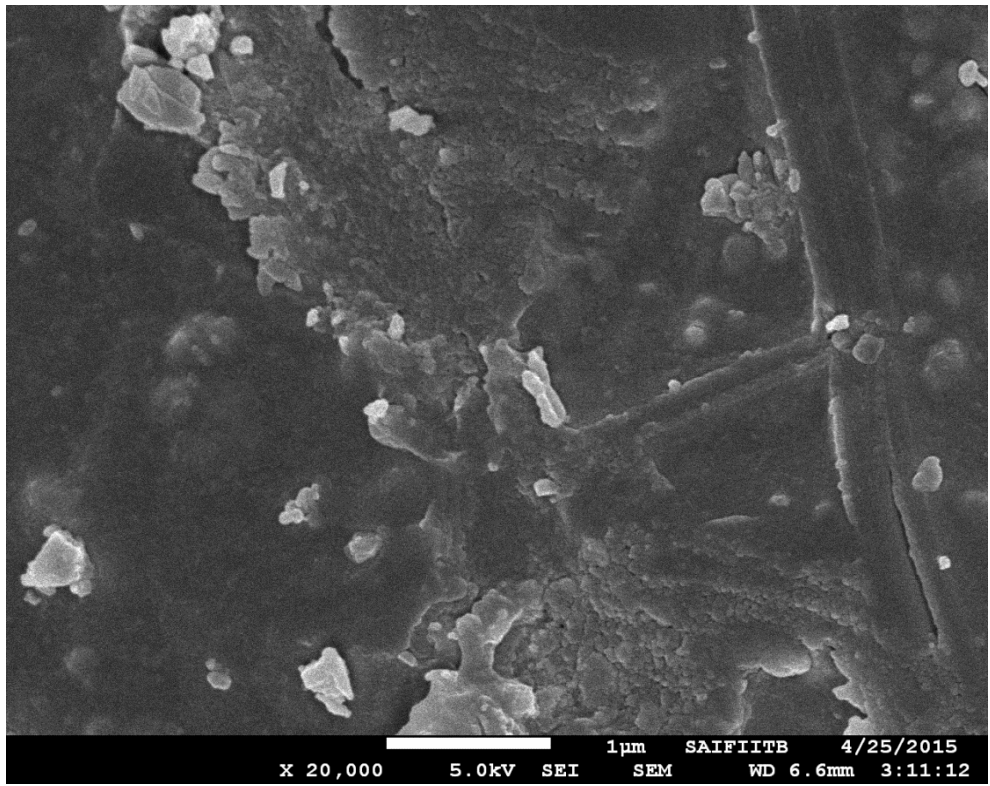

Fig 3.10: Aspirin loading of ratio 5:1 as per Table 4.2

### 3.1.2.2 STRONTIUM RANELATE

The drug content was evaluated for all the concentrations and it was observed that the nanoparticles of only CaCO$_3$ that showed inter-particle spaces and crevices were now appeared filled up. The drug thus is entrapped within or between the nanoparticles. These results were mainly attributed to aspirin powders increasing the charge density. Various solvent systems like distilled water, NaCl solution and mixture of NaCl : water were tried and the NaCl : water proved to give the best loading of drug. Maximum loading obtained in this work is also 50%. The water is used because it helps to keep the concentration of NaCl considerably less, so that there is more loading of drug particles and not NaCl. This is why NaCl must be dissolved completely in distilled water, before trying to dissolve SR in it. More volume of water is mainly to reduce NaCl concentration for better release study in LCMS so as to not saturate or block the column. Addition of water makes the process of evaporation slower to a good extent causing lesser harm to the nanoparticle's structures. Below are the FE-SEM images that show increase in the drug loading with an increase in the drug concentration while keeping the quantity of calcium carbonate powder same for all formulations.
It can be seen that the sample can be seen getting denser with an increase in concentrations, although, the first two concentrations of SR show almost similar images.

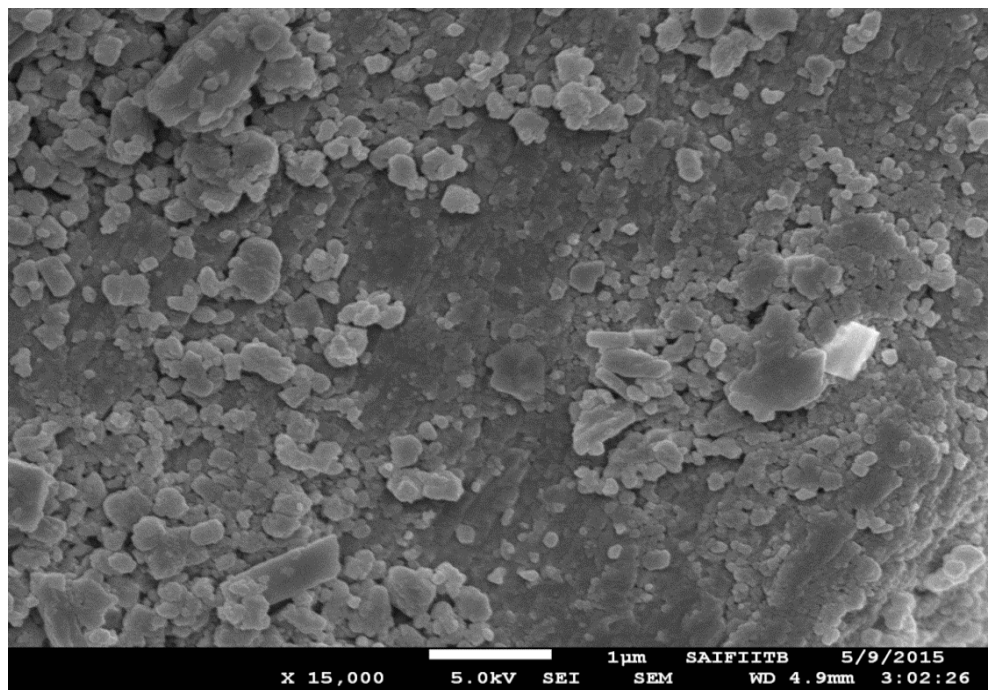

Fig 3.11: SR loading for ratio 1:20 as per Table 4.3

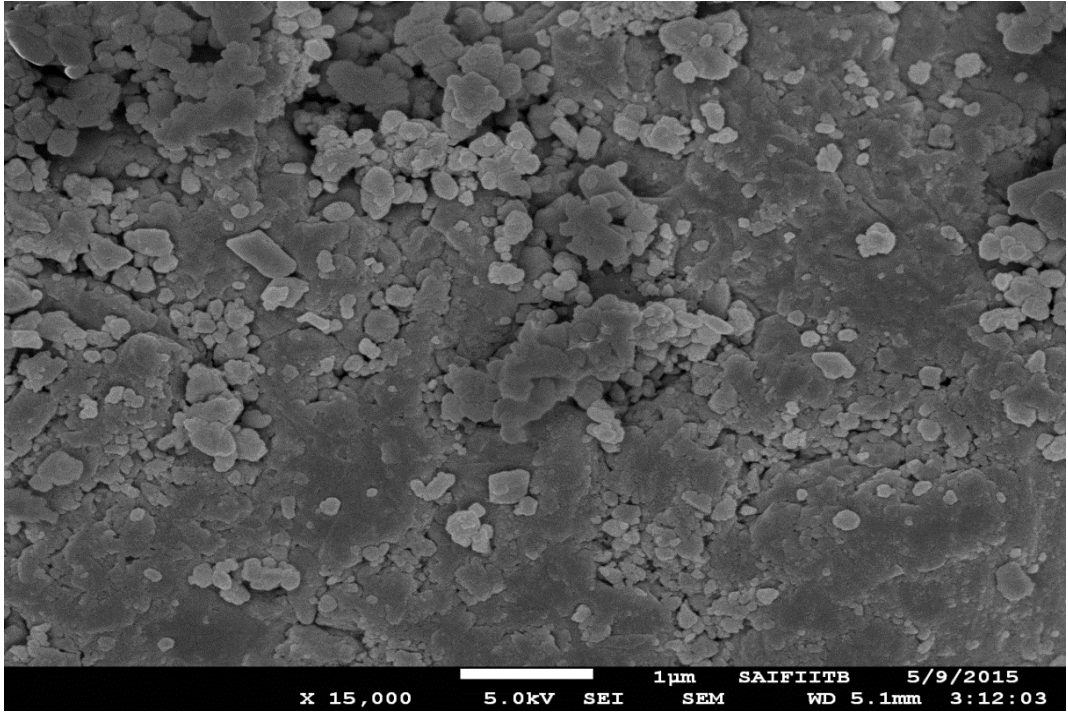

Fig 3.12: SR loading for ratio 1:4 as per Table 4.3

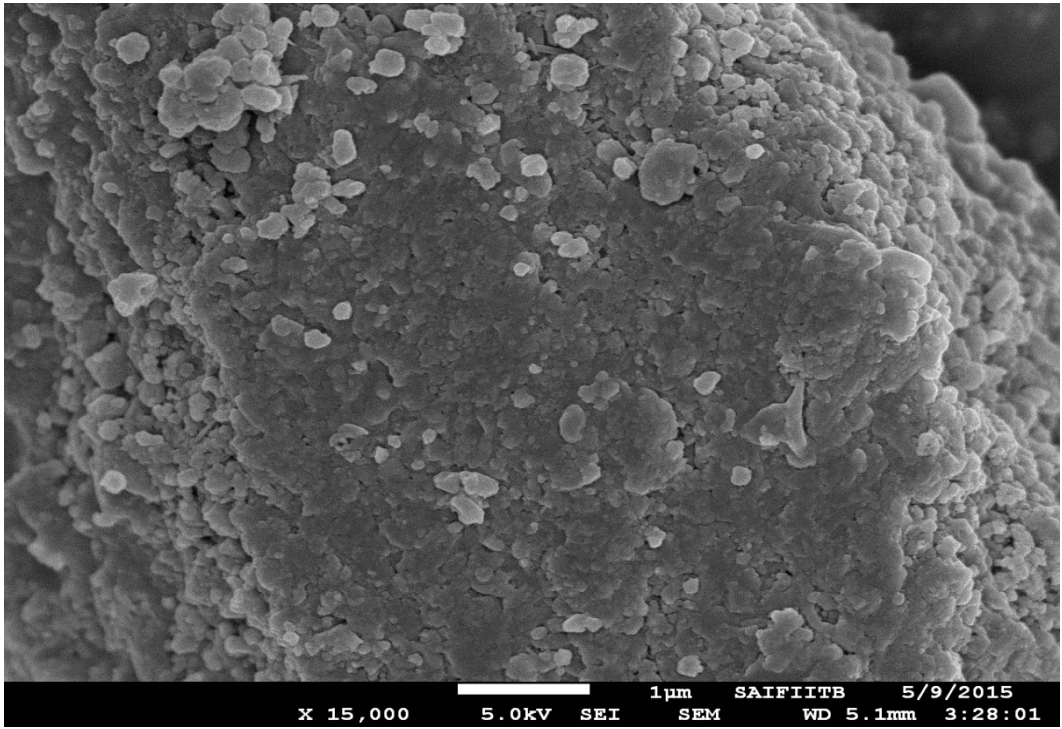

Fig 3.13: SR loading for ratio 1:2 as per Table 4.3

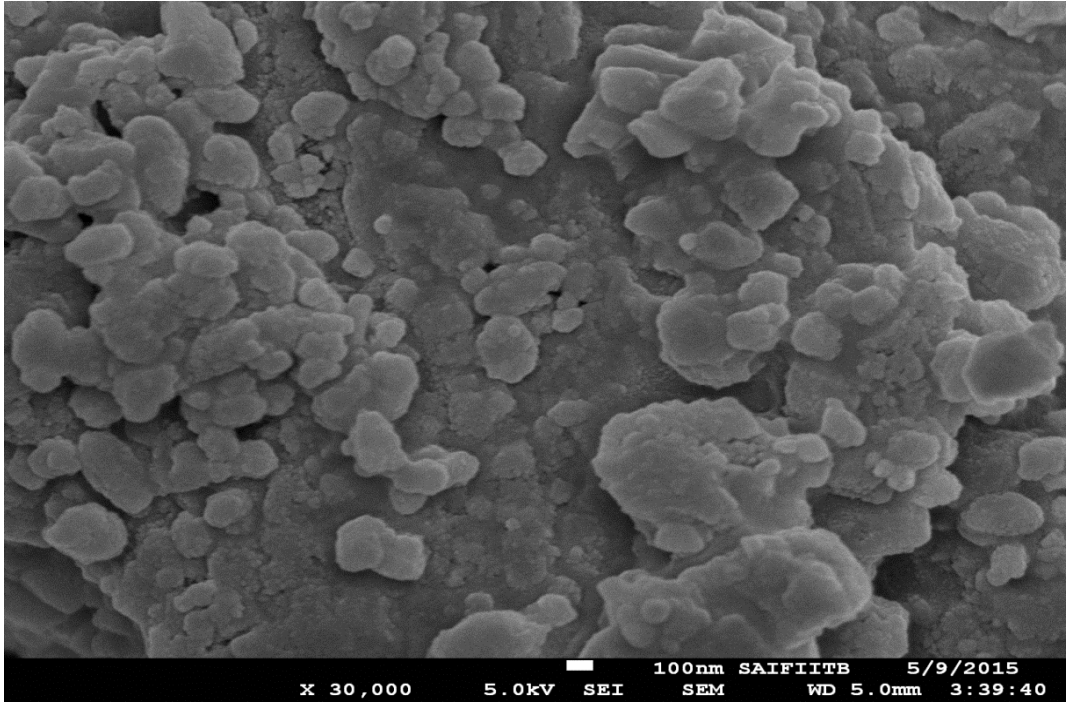

Fig 3.14: SR loading for ratio 1:1 as per Table 4.3

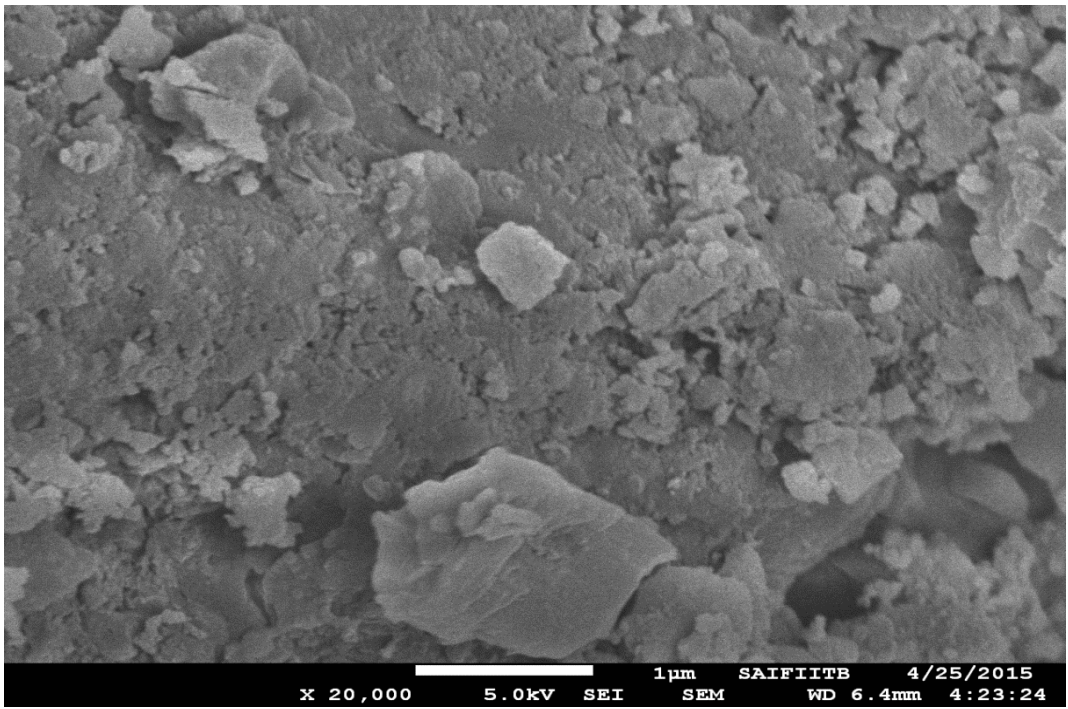

Fig 3.15: SR loading for ratio 2:1 as per Table 4.3

## 3.1.3 DRUG RELEASE STUDIES

### 3.1.3.1 ASPIRIN

The drug release for aspirin was done by HR-LCMS technique. A main objective of using porous carriers as a drug delivery system is to improve dissolution rate. Dissolution experimental methods have shown that drug loading into porous nanoparticles can accelerate drug release. The faster drug dissolution of the loaded drug was attributed to the large surface area of the calcium carbonate nanoparticles which are thus loaded efficiently. Increased crystallization pressure in pores leads to locally increased solubility and might be an additional mechanism for faster drug release. Drug release from drug-loaded calcium carbonate was investigated without specific formulation strategies to eliminate influences altering the dissolution rate, and to have direct information on the performance of drug-loaded particles.

| Conc. (µg/ml) ↓ | 1st hour | 2nd hour | 3rd hour | 4th hour | 5th hour | Last hour (after 19 hrs) |
|---|---|---|---|---|---|---|
| | Peak Areas | | | | | |
| 500 | 1304054 | 1110467 | 1236587 | 9911476 | 1207767 | 1286887 |
| 1000 | 1674158 | 1414768 | 1314158 | 1360398 | 1471738 | 5406717 |
| 2000 | 1329527 | 2057347 | 2192447 | 2536127 | 2440587 | 1163997 |
| 5000 | 7416827 | 9145467 | 9363317 | 1016338 | 9608017 | 6097227 |
| 10000 | 8169697 | 9484467 | 9488767 | 1051198 | 1024598 | 5774576 |

Table 3.1: High-Resolution LCMS areas according to concentration and time

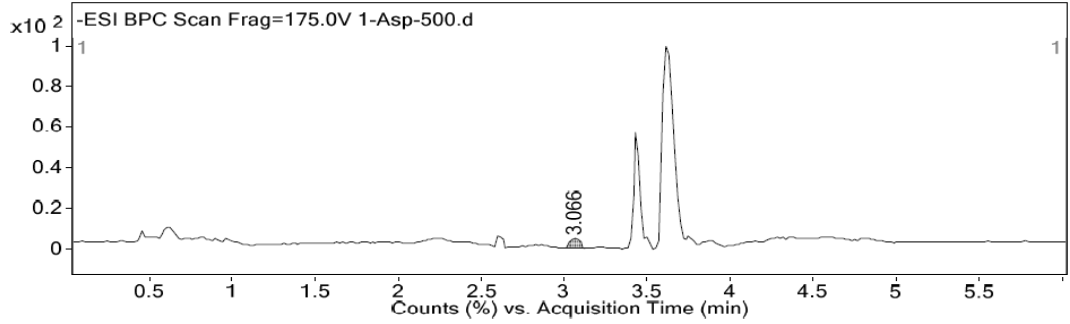
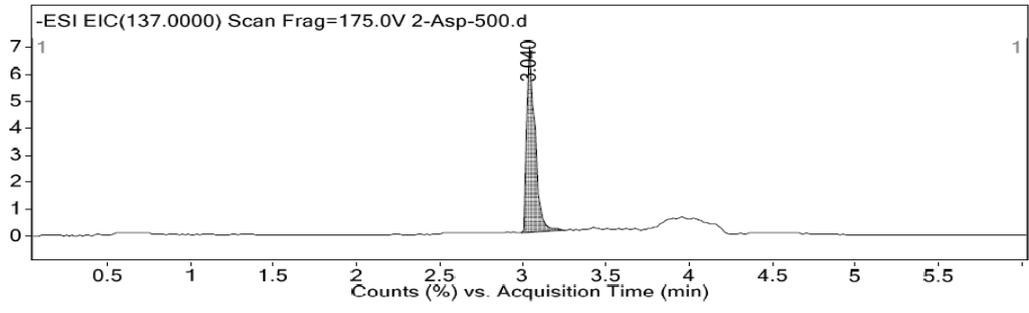
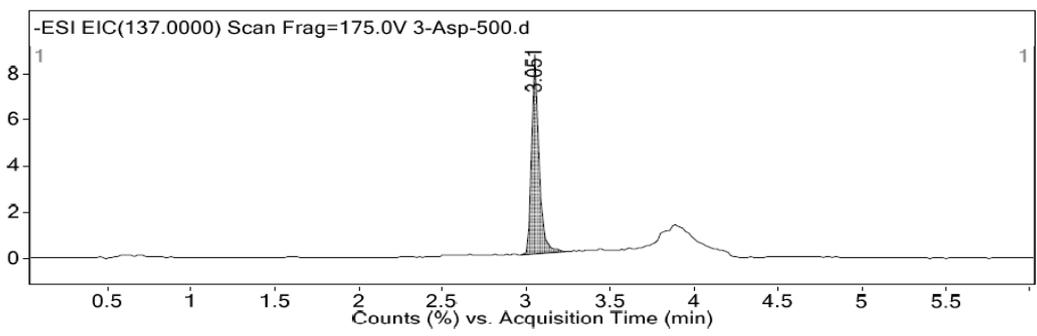
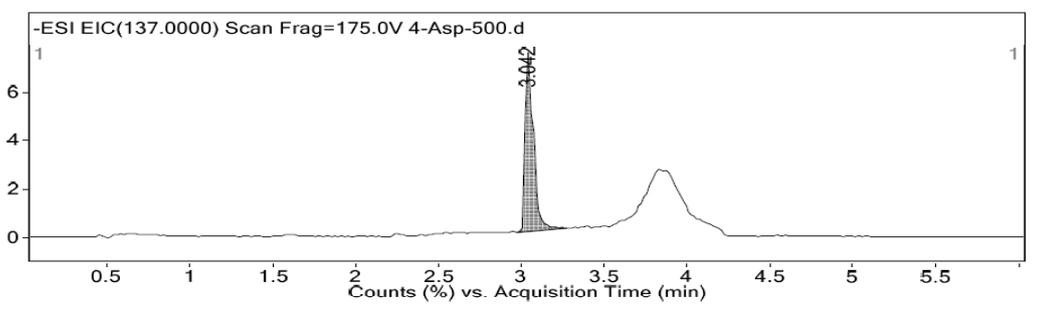

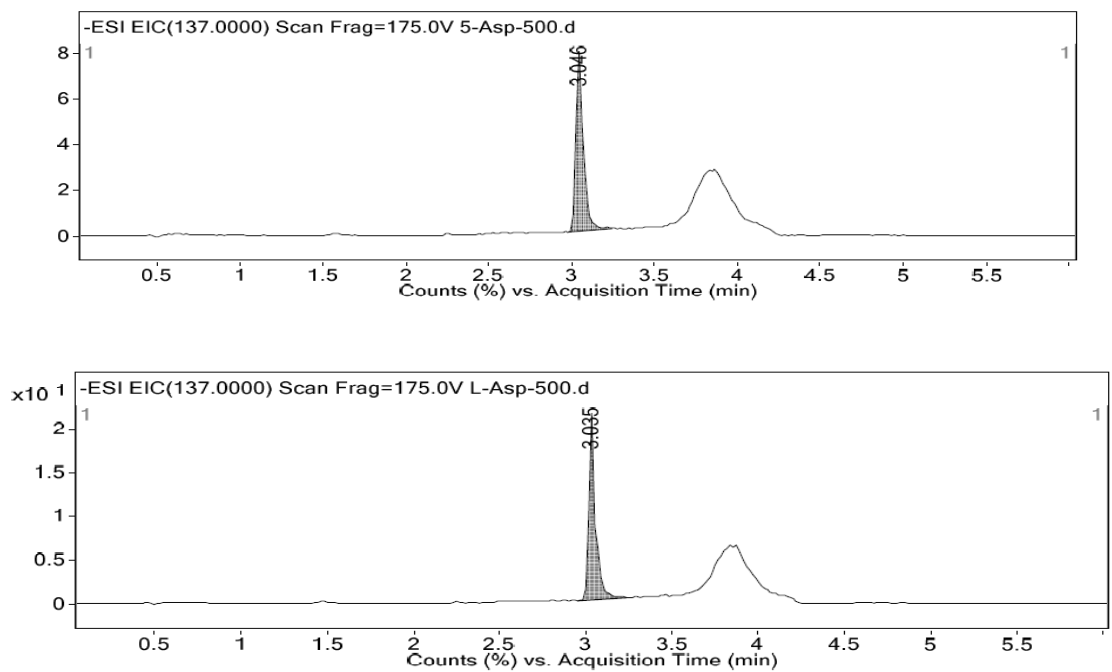

Fig 3.16: The results for the concentration 500 µg/ml show that aspirin released showed fast release from the 2nd hour till the last hour. Although considering the peak areas, it can be seen that the release shows exponential increase at the beginning of second hour and reduces as it moves towards the last hour.

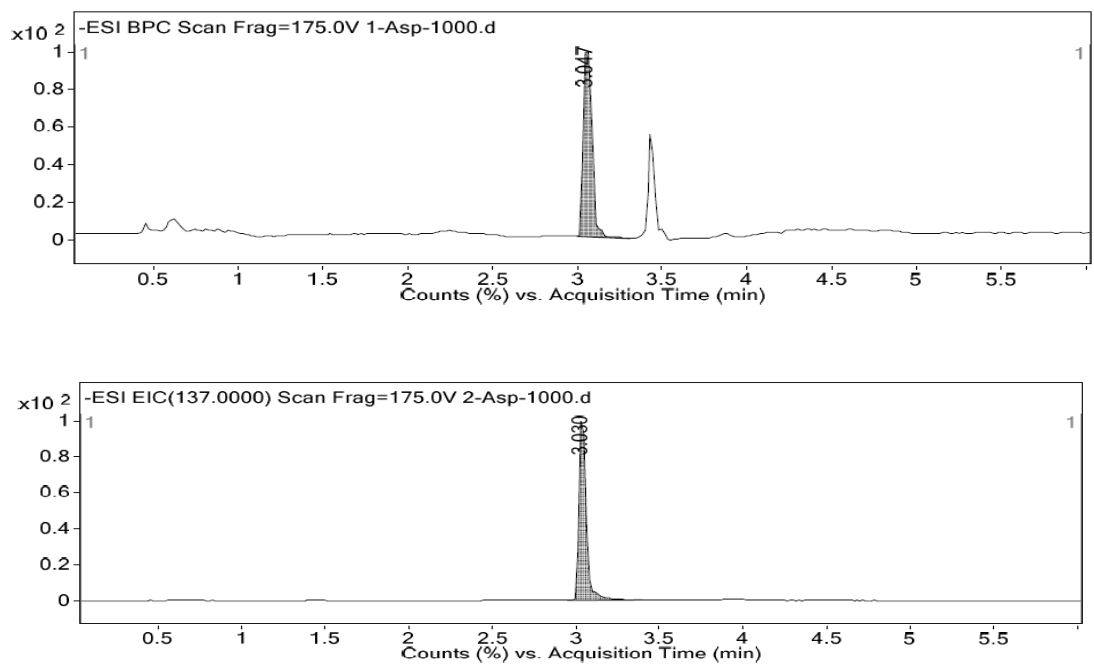

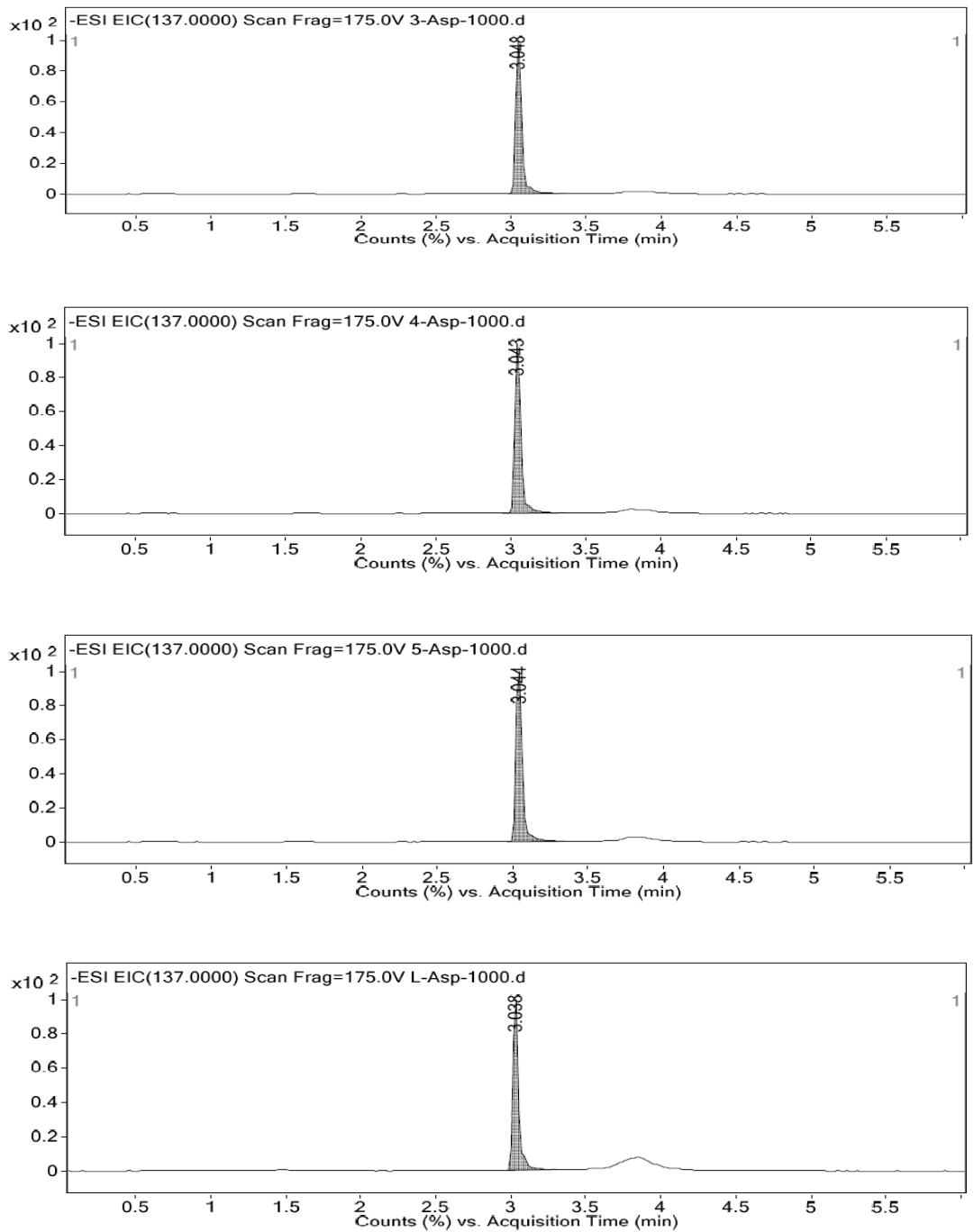

Fig 3.17: The HR-LCMS results for the concentration 1000 µg/ml show that aspirin released at a rapid release right from the first hour till the last hour. Although considering the peak areas, it can be seen that the release shows exponential increase at the beginning of second hour and reduces as it moves towards the last hour.

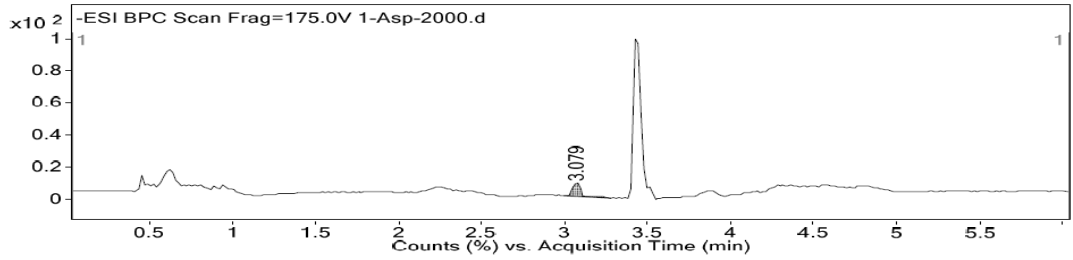
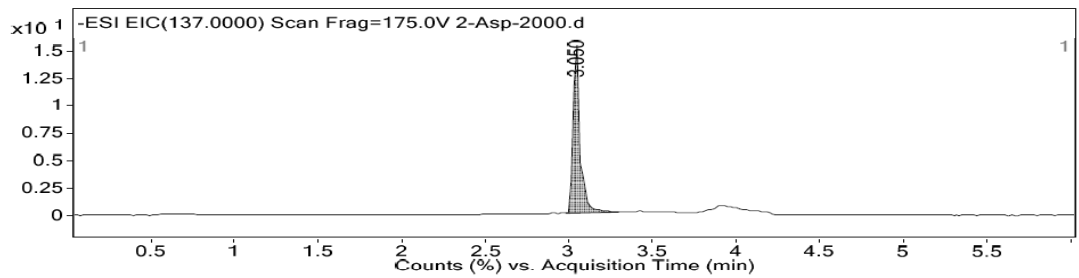
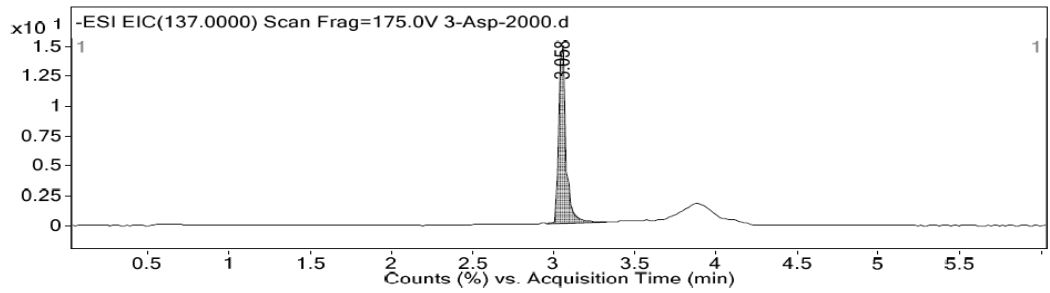
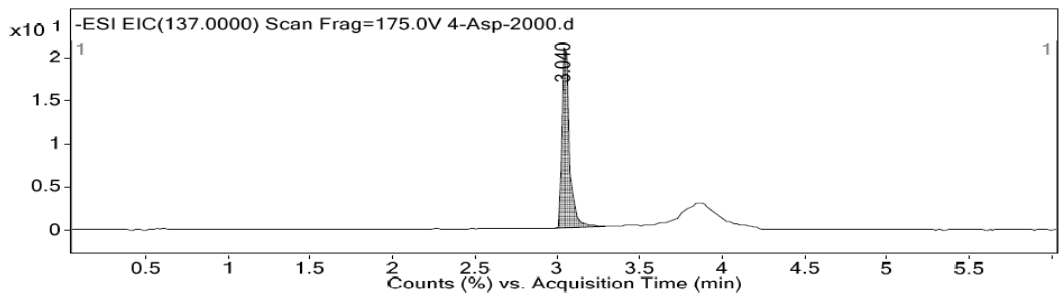

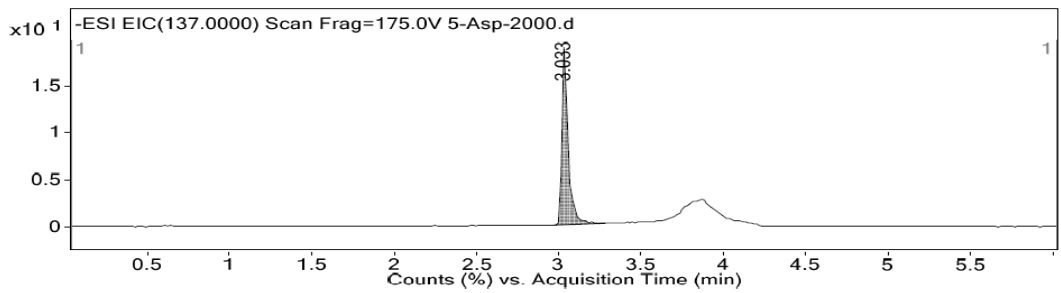

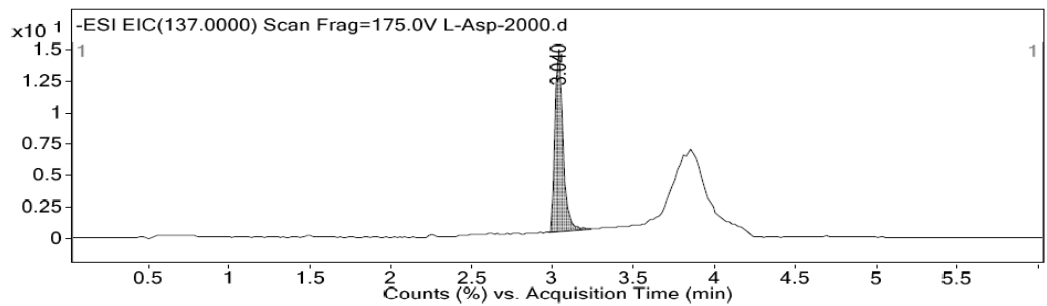

Fig 3.18: The HR-LCMS results for the concentration 2000 µg/ml show that aspirin released at a rapid release right from the second hour till the last hour. Although considering the peak areas, it can be seen that the release shows exponential increase at the beginning of second hour and reduces as it moves towards the last hour.

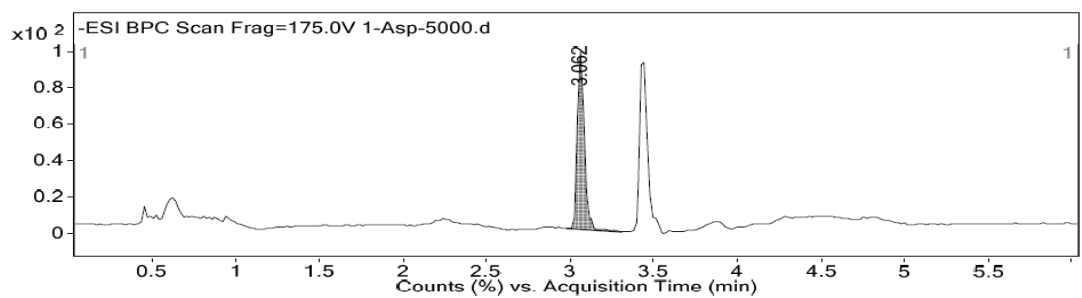

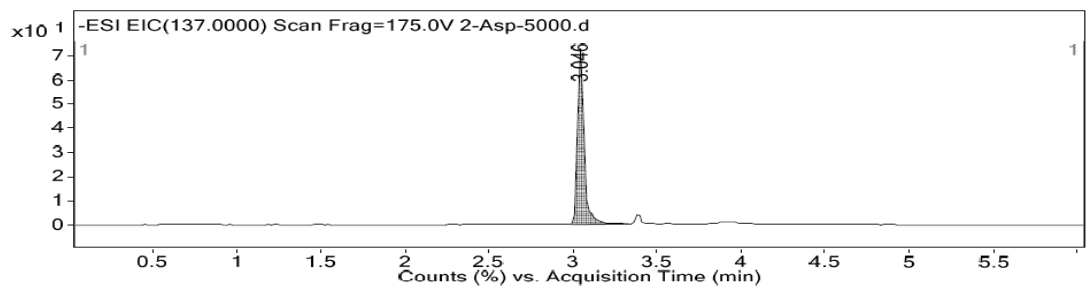

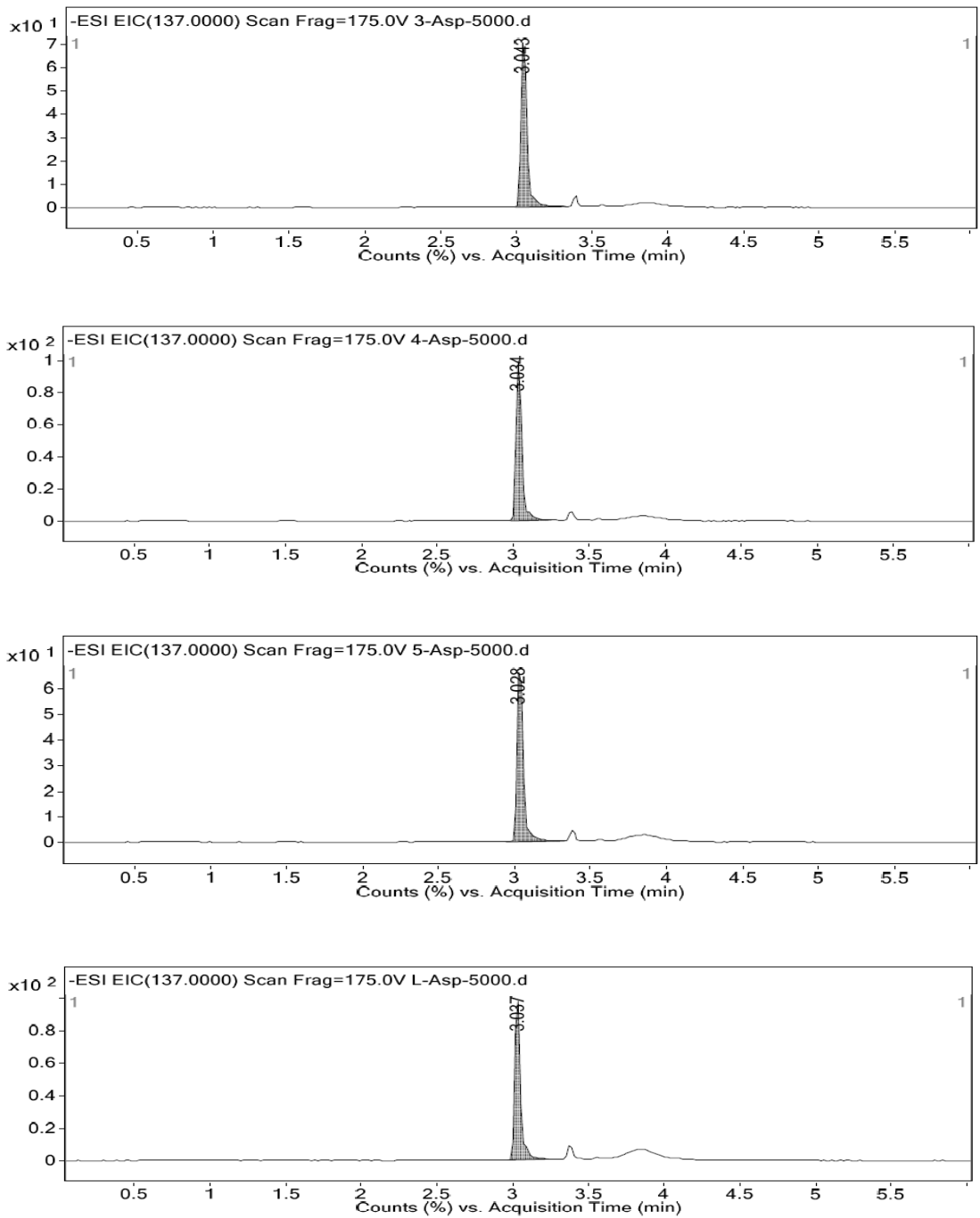

Fig 3.19: The HR-LCMS results for the concentration 5000 µg/ml show that aspirin released at a rapid release right from the first hour till the last hour. Although considering the peak areas, it can be seen that the release shows exponential increase at the beginning of second hour and reduces as it moves towards the last hour.

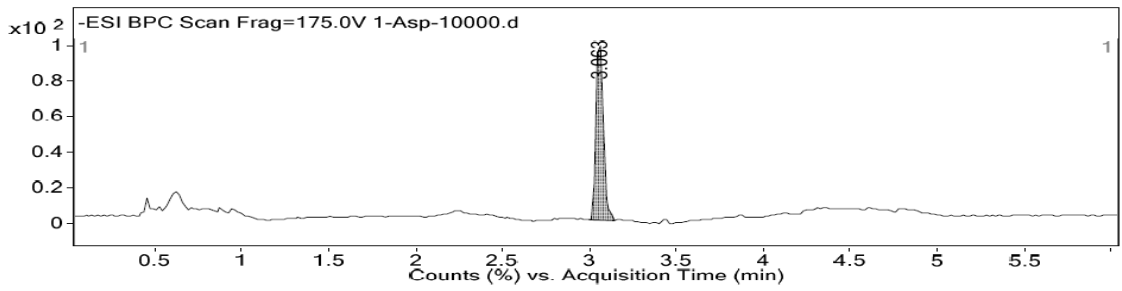
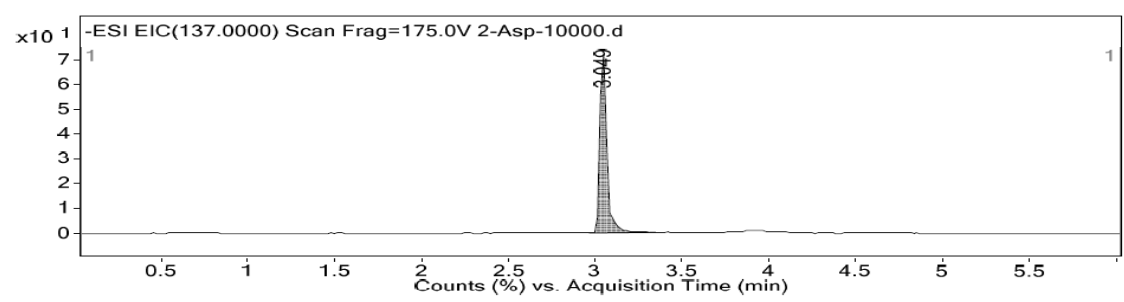
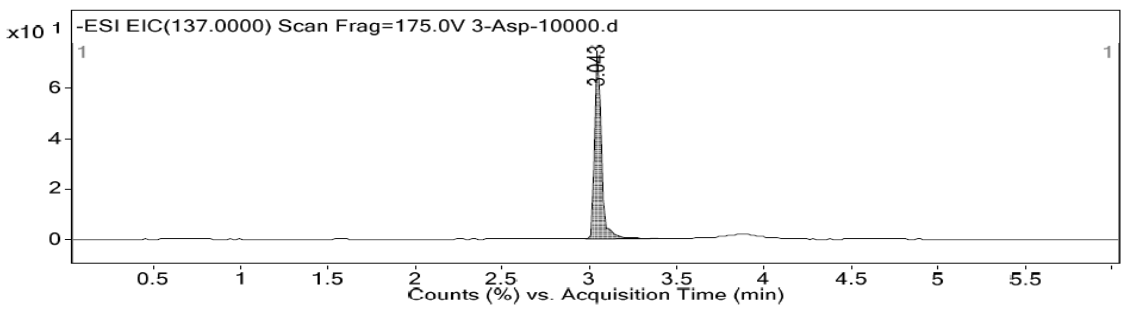
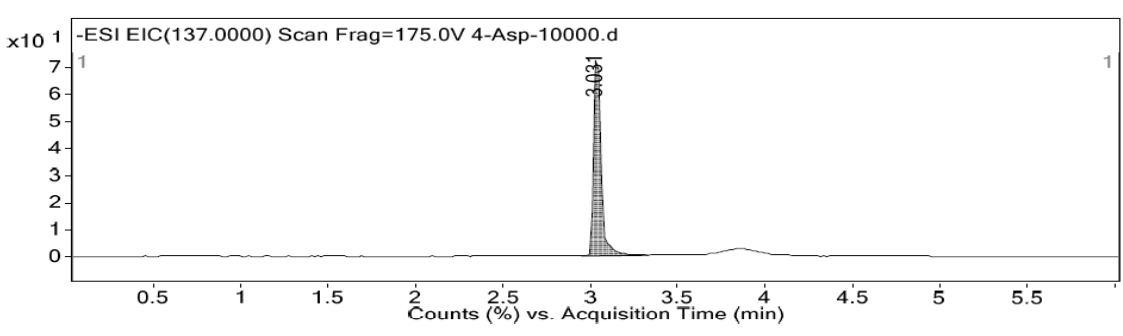
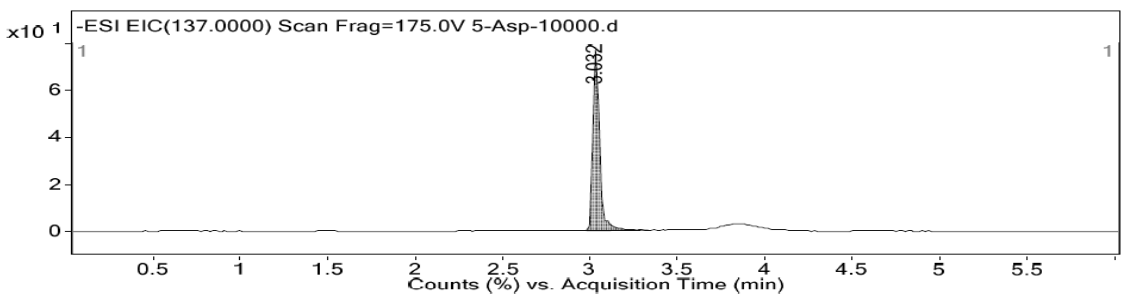

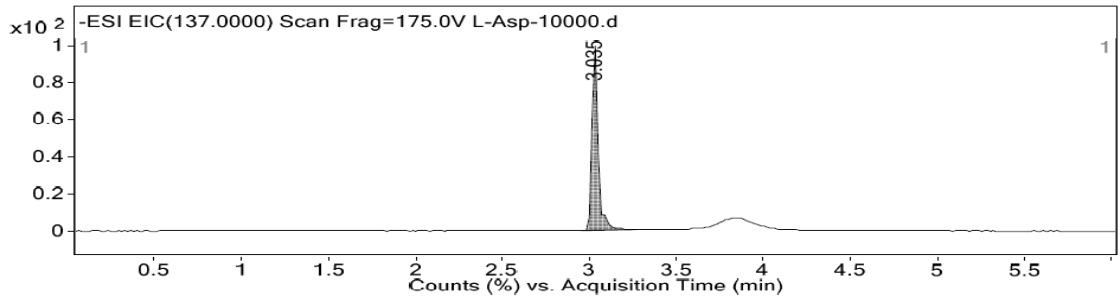

Fig 3.20: The HR-LCMS results for the concentration 10,000 µg/ml show that aspirin released at a rapid release right from the first hour till the last hour. But compared to first 5 hours, the release was lesser. Although considering the peak areas, it can be seen that the release shows exponential increase at the beginning of second hour and reduces as it moves towards the last hour.

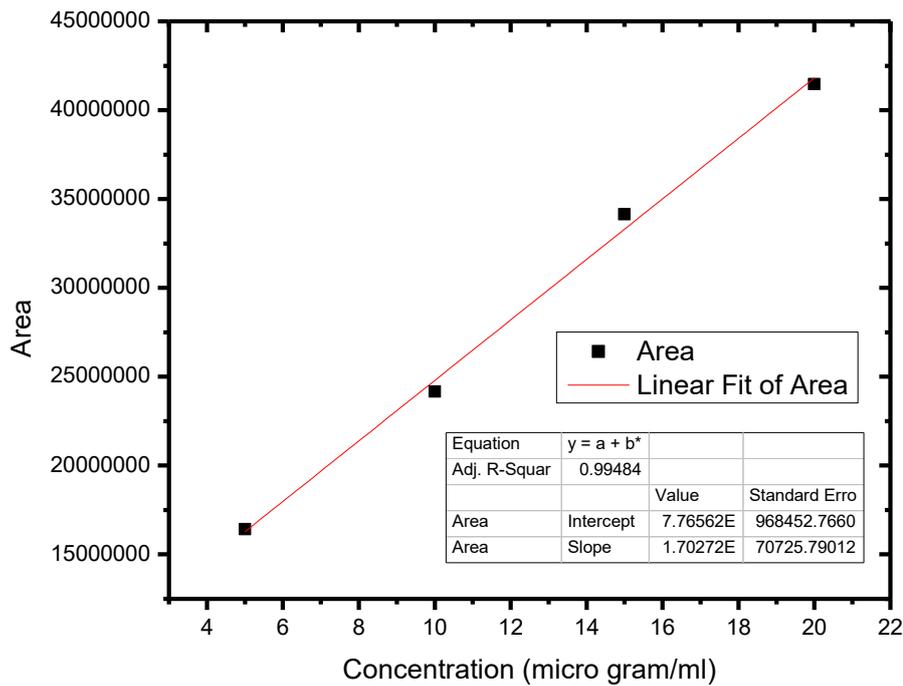

Fig 3.21: The calibration curve of linearity for Aspirin.

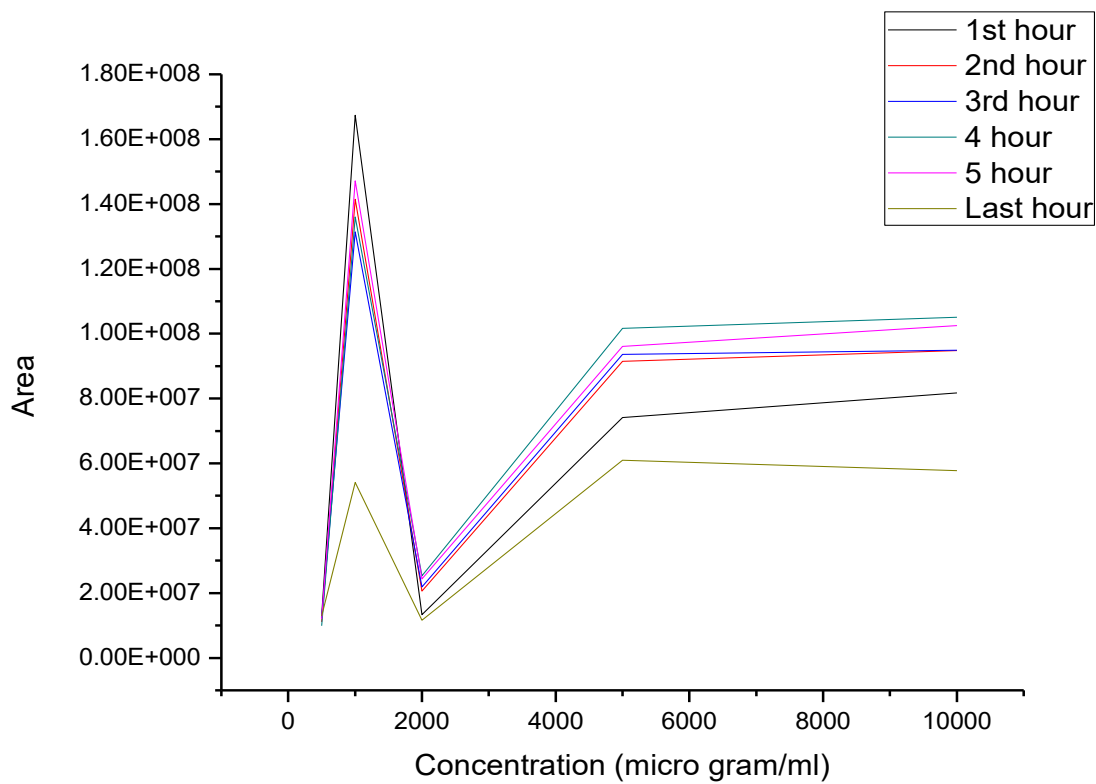

Fig 3.22: The graphical representation of the release of all the concentrations.

| Concentrations (µg/ml) | Drug : CaCO$_3$ |
|---|---|
| 500 | 1 : 16 |
| 1000 | 1 : 5 |
| 2000 | 1 : 1 |
| 5000 | 2.5 : 1 |
| 10000 | 5 : 1 |

Table 3.2: The ratios of aspirin drug to calcium carbonate for loading

### 3.1.3.2  STRONTIUM RANELATE

Different concentrations of the drug (SR) were prepared from the stock solution, ranging from 5 µg-100 µg/ml. Optical density measurements were taken at 323 nm by the UV-Vis spectrophotometer and a calibration curve was plotted

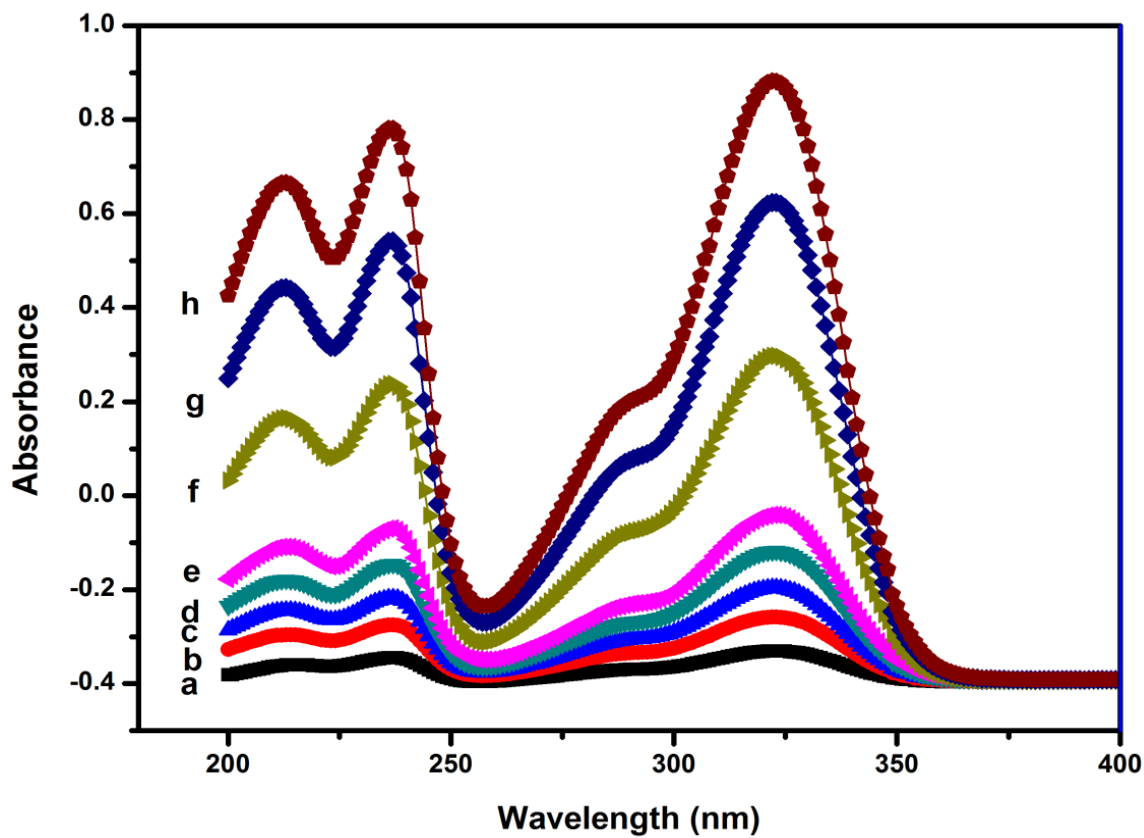

Fig 3.23: Overlay spectra of linearity (5-100 µg/ml) of Strontium ranelate

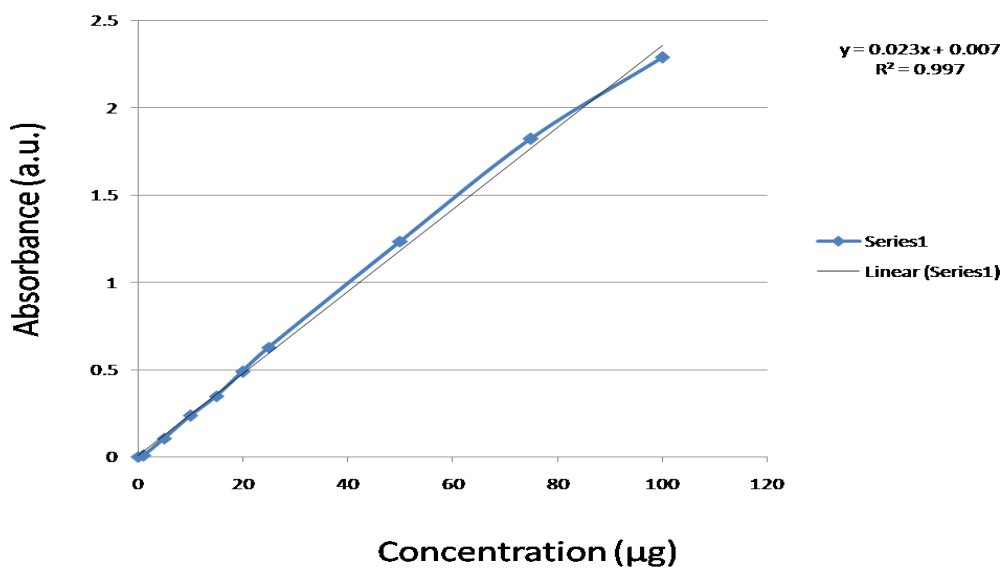

Fig 3.24: Calibration Curve of Strontium ranelate

| Drug : CaCO$_3$ | Drug: CaCO$_3$ ratio | Entrapment Efficiency (%) |
|---|---|---|
| 5:100 | 1:20 | 17.88 (approx.18%) |
| 25:100 | 1:4 | 24.86 (approx.25%) |
| 50:100 | 1:2 | 33.56 |
| 100:100 | 1:1 | 37 |
| 200:100 | 2:1 | 39 |

Table 3.3: Loading efficiency of SR

Specific concentrations of the drug (SR) and calcium carbonate were prepared. Drug loaded calcium carbonate particles were filtered and the filtrates were subjected for UV-Vis studies. Using UV-Vis spectroscopy the concentration values were determined. Using the values of the intercept and the slope, concentrations of the drug entrapped into the calcium carbonate were calculated. We can observe that the maximum entrapment is for the 2:1 ratio mixture. Further optimization experiments would be conducted in order to improve the loading efficiency of the system.

# CHAPTER 4

# CONCLUSION AND FUTURE SCOPE

# CONCLUSION

Contrary to common belief, we demonstrated that drug loading by solvent evaporation is simple, fast, and efficient. In particular, the solvent-evaporation method offers three advantages over the widely used impregnation method: the resulting drug loading can be precisely determined in advance, a high drug loading can be achieved in a short time, and the loading efficiency is good up to the loading capacity of calcium carbonate or other porous carriers, so there is also no waste of expensive drugs). The combination of qualitative FE-SEM analysis and HPLC quantification was sufficient to estimate both loading efficiency and loading capacity. Nevertheless, drug loading up to 50% with good release was achieved. Proof of feasibility of the solvent-evaporation method was mainly attributed to the unique properties of the model carrier used in this study. Drug-loading capacity could possibly be increased by slowing evaporation even more, thus allowing enough time for diffusion of drug into the core of the particles. Dissolution experiments proved that drug release can be accelerated by loading poorly water-soluble drugs into nanoparticles by the solvent-evaporation method. In contrast to the impregnation method based on adsorption, solvent evaporation led to a low fraction of amorphous drug after drug loading. The faster drug release can be explained by surface enlargement of the drug which may be caused due to heat during drying even though less and locally increased drug solubility.

# FUTURE SCOPE

Drug-loaded calcium carbonate may be used in capsules or tablets for oral delivery. Possible applications include the formulation of poorly soluble compounds to accelerate the dissolution. Sustained release formulations of highly soluble drugs can be obtained by coating of loaded particles. Technical advantages include the formulation of drugs which are administered at low doses, or the preparation of orally dispersible tablets. These findings are promising for the future development of carrier-based drug delivery systems.

# REFERENCES

1. Samuel A. Wickline, et al., Nanotechnology for Molecular Imaging and Targeted Therapy. *Circulation,* 107: 1092-1095. (2003)

2. Small is the new trend in the medical industry, *Digi News*, from http://www.digimouth.com/news/small-is-the-new-big-in-the-medical-world.html, (2011)

3. Frank X. Gu, et al., Targeted nanoparticles for cancer therapy. *Nano Today*, Volume 2, Issue 3, Pages 14-21. (2007)

4. Michael Goldberg, et al., Nanostructured materials for applications in drug delivery and tissue engineering. *Journal of Biomaterials Science Polymer Edition*, 18(3): 241–268. (2007)

5. How nanotechnology is shaping stem cell research, *The Guardian*, from http://www.theguardian.com/nanotechnology-world/nanotechnology-shaping-stem-cell-research (2012)

6. Cuenca AG., et al., Emerging implications of nanotechnology on cancer diagnostics and therapeutics. *Cancer*, 107(3):459-66, (2006)

7. Stephen Mahler, et al., Advances in drug delivery. *Journal of Chemical Technology and Biotechnology*, DOI: 10.1002/jctb.4689 (2015)

8. Murugan Ramalingam, et al., Tissue Engineering and Regenerative medicine: A Nano approach. CRC Press, 454-458, (2012)

9. Aspirin, *NHS Choices*, from http://www.nhs.uk/Conditions/Anti-platelets-aspirin-low-dose-/Pages/Introduction.aspx

10. What is aspirin? What is aspirin for? from http://www.medicalnewstoday.com/articles/161255.php

11. Aspirin use during pregnancy. *Drugs.com.*, from http://www.drugs.com/pregnancy/aspirin.html

12. 'Hands-on' self-test materials for organic structure solution from powder diffraction data, *International union of crystallography*, http://www.iucr.org/resources/commissions/powder-diffraction/projects/organic-datasets (2009)

13. A. Semalty, et al., Development and characterization of aspirin-phospholipid complex form improved drug delivery. *International Journal of Pharmaceutical Sciences and Nanotechnology*, Volume 3, Issue 2, (2010)

14. A. Salehi, et al., Simultaneous UV-VIS spectrophotometric determination of aspirin and methocarbamol in tablets. *Research in Pharmaceutical Sciences*, 7(5), (2012)

15. Mancini Marisabel Mourelle, et al., New strontium salts, synthesis and use thereof in the treatment of osteoporosis. *EP 2 530 068 A1* (2012)

16. Find a vitamin or supplement strontium. *WebMD - Therapeutic Research Faculty*, from http://www.webmd.com/vitamins-supplements/ingredientmono-1077-strontium.aspx?activeingredientid=1077&activeingredientname=strontium (2009)

17. Reginster JY., Strontium ranelate in osteoporosis. *Current Pharmaceutical Design*, 8(21):1907-16, (2002)

18. Abdullahi Shafiu Kamba, et al., Synthesis and Characterisation of Calcium Carbonate Aragonite Nanocrystals from Cockle Shell Powder (Anadara granosa). *Journal of Nanomaterials*, Article ID 398357, 9 pages (2013)

19. Daniel Preisig, et al., Drug loading into porous calcium carbonate microparticles by solvent evaporation. *European Journal of Pharmaceutics and Biopharmaceutics*, Volume 87, Issue 3, Pages 548–558 (2014)

20. Y. Ueno, et al., Drug-incorporating calcium carbonate nanoparticles for a new delivery system. *Journal of Controlled Release*, Volume 103, Issue 1, Pages 93–98, (2005)

21. SEM/EDS - Scanning Electron Microscopy with X-ray microanalysis, from http://wings.buffalo.edu/faculty/research/scic/sem-eds.html

# APPENDIX

1.  **HR-LCMS SOFTWARE**

    Agilent Technologies

    Mass hunter workstation software. LCMS data acquisition for 6200 series TOF/6500 series Q-TOF.

    Version B.05.01

    Build 5.01.5125.1

2.  **X-RAY DIFFRACTION**

    Philips powder diffractometer PW3040/60 X'pert pro

3.  **UV-VIS SPECTROPHOTOMETRY**

    Agilent Technologies

    Cary 100 series

4.  **SCANNING ELECTRON MICROSCOPY**

    JSM – 7600 F

    Software: PC-SEM

    Sputtering machine: JEOL JFC – 1600 Auto fine coater